\documentclass[twocolumn,tighten,referee]{aastex63}
\turnoffeditone
\turnoffedittwo
\usepackage[T2A]{fontenc}           % кодировка
\usepackage[utf8]{inputenc}         % кодировка исходного текста
\usepackage[english]{babel}
\usepackage{graphicx}  % Для вставки рисунков
\graphicspath{{img/}} % Пути к изображениям
\usepackage{xspace}  % Для вставки рисунков

\usepackage{amsmath}
\newcommand{\snref}{SN\,Refsdal\xspace}

\newcommand{\sna}{{SN~1987A\xspace}}
\definecolor{darkgreen}{rgb}{0.,0.7,0.}

\newcommand{\stella}{{\sc Stella}\xspace}

\newcommand{\rsun}{\ensuremath{R_\odot}}
\newcommand{\msun}{\ensuremath{M_\odot}}

\newcommand\nifsx{$^{56}$Ni\xspace}

\shorttitle{SN Refsdal}
\shortauthors{Baklanov et al.}
\begin{document}
\title{Strongly lensed SN Refsdal: refining time delays based on the supernova explosion models}

\correspondingauthor{Petr Baklanov}
\email{petr.baklanov@itep.ru}

\author{Petr Baklanov}
\affiliation{NRC "Kurchatov Institute" - ITEP 117218 Moscow, Russia}
\affiliation{Space Research Institute (IKI), Russian Academy of Sciences, Profsoyuznaya 84/32, 117997 Moscow, Russia}
\affiliation{National Research Nuclear University MEPhI, Kashirskoe sh. 31, Moscow 115409, Russia}

\author{Natalia Lyskova}
%\affiliation{National Research University Higher School of Economics, Myasnitskaya str. 20, Moscow 101000, Russia}
\affiliation{Space Research Institute (IKI), Russian Academy of Sciences, Profsoyuznaya 84/32, 117997 Moscow, Russia}
\affiliation{P.N.Lebedev Physical Institute, Leninskiy prospect 53, Moscow 119991, Russia}

\author{Sergei Blinnikov}
\affiliation{NRC "Kurchatov Institute" - ITEP 117218 Moscow, Russia}
%\affiliation{M.V. Lomonosov Moscow State University, Sternberg Astronomical Institute, Universitetsky~pr.,~13,~Moscow, 119234, Russia}
%\affiliation{Institute for Theoretical and Experimental Physics (Kurchatov Inst., ITEP), 117218 Moscow, Russia}
\affiliation{Kavli Institute for the Physics and Mathematics of the Universe (WPI), The University of Tokyo, Kashiwa, Chiba 277-8583,  Japan}

%\author{Surhud More}
%\affiliation{Kavli Institute for the Physics and Mathematics of the Universe (WPI), The University of Tokyo, Kashiwa, 277-8583,  Japan}

\author{Ken'ichi Nomoto}
\affiliation{Kavli Institute for the Physics and Mathematics of the Universe (WPI), The University of Tokyo, Kashiwa, Chiba 277-8583,  Japan}

\begin{abstract}
We explore the properties of supernova (SN) ``Refsdal'' - the first discovered gravitationally lensed SN with multiple images.
A large magnification provided by the galactic-scale lens, augmented by the cluster lens, gave us a unique opportunity to perform a detailed modelling of a distant SN at $z \simeq 1.5$.
We present results of radiation hydrodynamics modelling of \snref.
According to our calculations, the \snref progenitor is likely to be a more massive and energetic version of \sna, i.e. a blue supergiant star \edit1{with the following parameters: the progenitor radius  $R_0 = (50 \pm 1) R_{\odot}$, the total mass $M_{tot}= (25\pm 2) M_{\odot}$, the radioactive \nifsx mass $M_{^{56}\mathrm{Ni}} = (0.26 \pm 0.05) \,M_{\odot}$, and the total energy release   $E_{burst}=(4.7 \pm 0.8)\times10^{51}$ erg. }
Reconstruction of  SN light curves allowed us to obtain  time delays and magnifications for the  images S2-S4 relative to S1 with higher accuracy than previous  \edit1{template-based estimates} of \cite{Rodney2016}. \edit2{The measured time delays are $\Delta t_{S2-S1} = 9.5^{+2.6}_{-2.7}  $ days, $\Delta t_{S3-S1} = 4.2^{+2.3}_{-2.3} $ days, and $\Delta t_{S4-S1} = 30^{+7.8}_{-8.2} $ days. The obtained magnification ratios are $\mu_{S2/S1}= 1.14 \pm  0.02$, $\mu_{S3/S1} = 1.01 \pm 0.02 $, and $\mu_{S4/S1} =  0.35\pm  0.02$.}
We estimate the Hubble constant $H_0 = 68.6^{+13.6}_{-9.7}$ km s$^{-1}$ Mpc$^{-1}$ via re-scaling  the time delays predicted by different lens models to match the values obtained in this work. 
With more photometric data on the fifth image SX, we will be able to further refine  the time delay and magnification estimates for SX and obtain competitive constraints on $H_0$.
%These measurements provide a direct way to infer  the Hubble constant and, in principle, other cosmological parameters.

\end{abstract}


\section{Introduction} 
\label{sec:intro}

Supernova explosions are among the most energetic and fascinating phenomena in the Universe.
Investigating these objects is essential not only for understanding the physics of stellar explosions, but also for studying properties of progenitor population, stellar evolution, nucleosynthesis, modeling chemical evolution of galaxies, origin of cosmic rays, to name a few. 
Throughout modern astrophysics, supernovae (SNe) have been also used to measure cosmological distances. 
Due to  high intrinsic brightness and ``standardizable'' light curves (LCs),
SN Ia are now routinely used to determine cosmological parameters. 
It was by using SNe Ia that  \cite{Riess1998} and \cite{Perlmutter1999} discovered an accelerated  expansion of the Universe. 
Observations  of  type  II  SNe  can  be  also used  to  determine distances to their host galaxies. 
Despite the fact that SNe II show a large variations in their observational properties (luminosities, durations, etc.), there are a number of methods to utilize observations of SN II for cosmological studies \citep[see, e.g., ][for a review]{Nugent2017}.  
For instance, the Expanding Photosphere Method (EPM) was proposed by  \cite{Kirshner.Kwan.1974}  to measure distances to the Type II plateau supernovae whose light curve is nearly flat  for $\sim$ 100 days and then suddenly drops off. 
The EPM has been successfully applied to nearby SN IIP \citep[e.g.][]{Tsvetkov2019} 
and more distant objects (up to $z \simeq 0.34$; \citealt{Gall2018}). 
Other techniques include the Spectral-fitting Expanding Atmosphere Method \citep[e.g.][]{Baron2004} for SN IIP and the Dense Shell Method \citep{Potashov2013, Baklanov2013} to 
measure distances to SN IIn supernovae.

One of the current frontiers in SNe research centers is constructing numerical models of SNe explosions, reliability of which can be determined from comparison with observational data. 
Such SN modelling requires high-quality photometric and spectroscopic data.
While for  super-luminous SNe such detailed information can be in principle obtained even at high redshifts $z>2$ \citep{Cooke2012},  this is not the case for type IIP SNe, which are typically observed up to $z \sim 0.4$ \citep{Nugent2017}. 
So far, hydrodynamical models of type IIP supernovae were constructed only for nearby objects.  

Recent discovery of gravitationally lensed supernovae with multiple images -- SN Refsdal \citep{Kelly2015_Discovery} and SN iPTF16geu  \citep{Goobar2017Sci} -- opens up a window  to the unexplored
high-redshift transient universe. 
Strongly lensed supernovae represent a class of objects unique both for astrophysics and cosmology. 
They make possible not only investigation of the properties of supernova progenitors (pre-supernovae) and their environments at high redshifts (a signal from which would not be detected in the absence of a lens), but they can be also used for cosmological studies. 
In a case of a variable source such as a supernova, light curves for different images are shifted in time relative to each other. 
By measuring these time delays between images, one can obtain an independent estimate \edit1{of} the Hubble constant (first suggested by \citealt{Refsdal1964}) and the dark energy equation of state \citep[e.g. ][]{Linder2011}.  
For certain types of SNe II and for SN Ia, the intrinsic luminosities 
 can be inferred independently of \edit1{gravitational} lensing. 
In such cases, the absolute lensing magnification 
can be constrained independently of a lens model, thus helping to break the degeneracy between the radial mass profile of a lens and the Hubble constant \citep[][]{Oguri.Kawano2003}. 
Indeed,  numerous studies of lensed quasars have convincingly shown  the Hubble constant value is sensitive to details of a lens model \citep[e.g. ][among others]{Kochanek2002,Larchenkova2011,Birrer2016,Wong2020} and assuming a  power-law density distribution (the simplest lens model) introduces  a bias in the determination of $H_0$ \citep[e.g. ][]{Xu2016MNRAS}. 
%(Kolatt  &Bartelmann 1998; Xu et al. 2016).

This paper is devoted to radiation-hydrodynamics modelling of the first discovered lensed supernova with multiple images -- SN Refsdal. 
\cite{Kelly2016} has already shown that the spectra and light curve of SN Refsdal are similar to  those of SN 1987A, a peculiar SN II in the Large Magellanic Cloud, and that the progenitor of SN Refsdal is most likely to be a blue supergiant star.
 As emphasized in \cite{Rodney2016}, none of existing light curve templates is able to capture all the features of  SN Refsdal light curve, thus making the task of modelling of SN Refsdal important. 
Moreover, SN Refsdal was located in the arm of a
spiral host galaxy at $z \simeq 1.5$, i.e. much farther away than any modelled Type II SNe so far. 
The construction of a physical model of the pre-supernova, which satisfies available photometric observations in different filters, should in principle allow one to determine time delays between images more accurately than it is done in \cite{Rodney2016} and to constrain the magnification factors. 
This information can serve as an independent test of  different lens models presented in the literature  \citep[see][for a compilation of lens models]{Treu2015} and/or used as an additional constraint to improve the lens model.  
The latter should lead to an improved precision in determining the Hubble constant and other cosmological parameters \citep[e.g., ][]{Grillo2018, Grillo2020}. 
The paper is organized as follows. 
In Section~\ref{sec:observations}, we list available observational data on SN Refsdal. 
Section~\ref{sec:snsim} gives a brief description of constructed hydrodynamical SN models, the best-fit model which matches all available observational data is described in Section~\ref{subsec:best_fit_model}  
Technical details on the fitting procedure are given in Appendix~\ref{adx:Bayesian}. 
We use the reconstructed SN Refsdal \edit1{light curve \deleted{(LC)}} to derive time delay and magnification ratios for all images in Section~\ref{sec:dt_and_mu}. 
With these estimates in hand, we obtain the most likely Hubble constant value in Section~\ref{sec:H0}. 
Finally, all the results of this work are summarized in Section~\ref{sec:conclusions}.

% % % % % % % % % % % % % % % % % % % % % % % % % % % % % % % % % % %
\section{Observations}
\label{sec:observations}
%\subsection{ Photometry and spectra}
A strongly lensed supernova was found in the MACS J1149.6+2223  galaxy cluster field on 10 November 2014 \citep{Kelly2015_Discovery}. The HST images revealed four resolved images of
the background SNe ($z=1.49$) arranged in an Einstein cross configuration around a massive elliptical galaxy ($z=0.54$) -- \edit1{a} MACS J1149.6+2223 cluster member.

To construct a hydrodynamic model \edit2{for} SN Refsdal, we use photometric data from \cite{Rodney2016} (their Table~4) obtained with HST using the Wide-Field Camera~3 (WFC3) with the infrared (IR) and UV-optical (UVIS) detectors, and the Advanced Camera for Surveys (ACS).

The dynamical properties of the envelope and characteristic expansion velocities can be obtained by investigating line profiles in the spectra of the supernova. 
Thanks to gravitational amplification of the SN Refsdal light, there are HST, Keck, and VLT X-shooter spectra \citep{Kelly2016}  available. 
Despite being noisy, these spectral observations give us constraints on how the velocity of the envelope was changing during the epoch of maximum light in the F160W band. We use the H$\alpha$ expansion velocity measurements from  \cite{Kelly2016} in Section \ref{sec:snsim} to constrain the model parameter space.

In the direction of SN Refsdal dust absorption in our Galaxy is insignificant,
\edit1{ with	$E(B-V)_{MW} = 0.02$ \citep{Lotz2017}}.
% and even in the U band does not exceed $ 0.1^m $ \citep{Lotz2017}.
This is not surprising, since for observations of distant objects, such as the galaxy  cluster MACS J1149.5+2223, it is natural to choose transparency windows in the Galaxy.
%There are no data of dust extinction for \snref \textbf{in} \cite{Kelly2016}.
%, thus we have not adjusted our theoretical  light curves to account for the host reddening.
%
\edit1{Unfortunately, there is no information on dust extinction for \snref, and theoretical light curves presented  in this work were not  corrected for    the host    reddening. One way to account for $E(B-V)_{host}$
would be to include it as a fit parameter. 
For example, \cite{Rodney2016} added magnitude shifts as free parameters for each photometric passband to account for any color difference between a template and \snref. 
However, such an approach seems not to be optimal for our task --- to model \snref light curves in various photometric bands self-consistently.}

%\textbf{
%	\cite{Rodney2016} have added a magnitude shifts as free parameters for each photometric passbands which allowed to take into account, among other things, the  dust extinction in the host.
%Using the host reddening $E(B-V)_{host}$ as a free parameter would allow us to better fit observed and theoretical light curves, 
%	but significantly decreased the constraints on theoretical light curves. 
%	We have decided not to add   $E(B-V)_{host}$  as free parameter in our fitting procedure.  	 
%}
%

% % % % % % % % % % % % % % % % % % % % % % % % % % % % % %
\section{Supernova simulation} \label{sec:snsim}
SN Refsdal light curves demonstrate the slow rise in brightness  to a broad peak. 
Combining this information with the analysis of H$\alpha$-emission and absorption features, 
\cite{Kelly2016} and \cite{Rodney2016}  have already shown that \snref is a peculiar 1987A-like SN.
SN~1987A, in its turn, is classified as a peculiar Type II Plateau SN with a progenitor 
being a blue supergiant,  rather than a red supergiant as for ordinary Type II-P SNe. 
SN~1987A has been intensively studied in recent decades (e.g., \cite{Utrobin2005}, 
for a review see \cite{1987aReview}). 

For the model calculation, we use the multi-group radiation-hydrodynamics numerical code \stella 
\citep[\edit1{STatic Eddington-factor Low-velocity Limit Approximation};][]{Blinnikov2004, Baklanov2005, Blinnikov2006}.
\stella allows one to construct synthetic light curves in various photometric bands and takes into account   available observational constraints on the expansion velocity 
(coming from the analyses of P~Cygni profiles),
 i.e. with \stella we can utilize  all the available  SN Refsdal observational data. 
\stella has been successfully used for a wide variety of supernova studies 
  including but not limited to super-luminous supernovae (SLSNe) and
  pulsational pair-instability supernovae (PPSNe), SNe\,Ia, SNe\,IIP  
  \citep{Woosley2007PPSN, Woosley2007snIa,Sahu2008,Tominaga2011,Baklanov2015,Sorokina2016}.

Note that \stella is 1D and does not allow \edit1{to take into account any changes} 
in a chemical composition \edit1{which are caused by} explosion-driven Rayleigh-Taylor instabilities and a shock wave passage through the SN shell. 
Moreover, for  \sna-like \edit1{SNe the} formation of a magnetar at the center of the SN shell is possible \citep{Chen2020}. 
That leads to an additional mixing of metals and complicates the distribution of chemical elements in the shell. 
Thus, we use `non-evolutionary' SN models and artificially reproduce details of the  evolutionary models as well as mixing during an explosion.

\begin{figure*}[htb!]
\centerline{
 \includegraphics[width=0.5\textwidth]{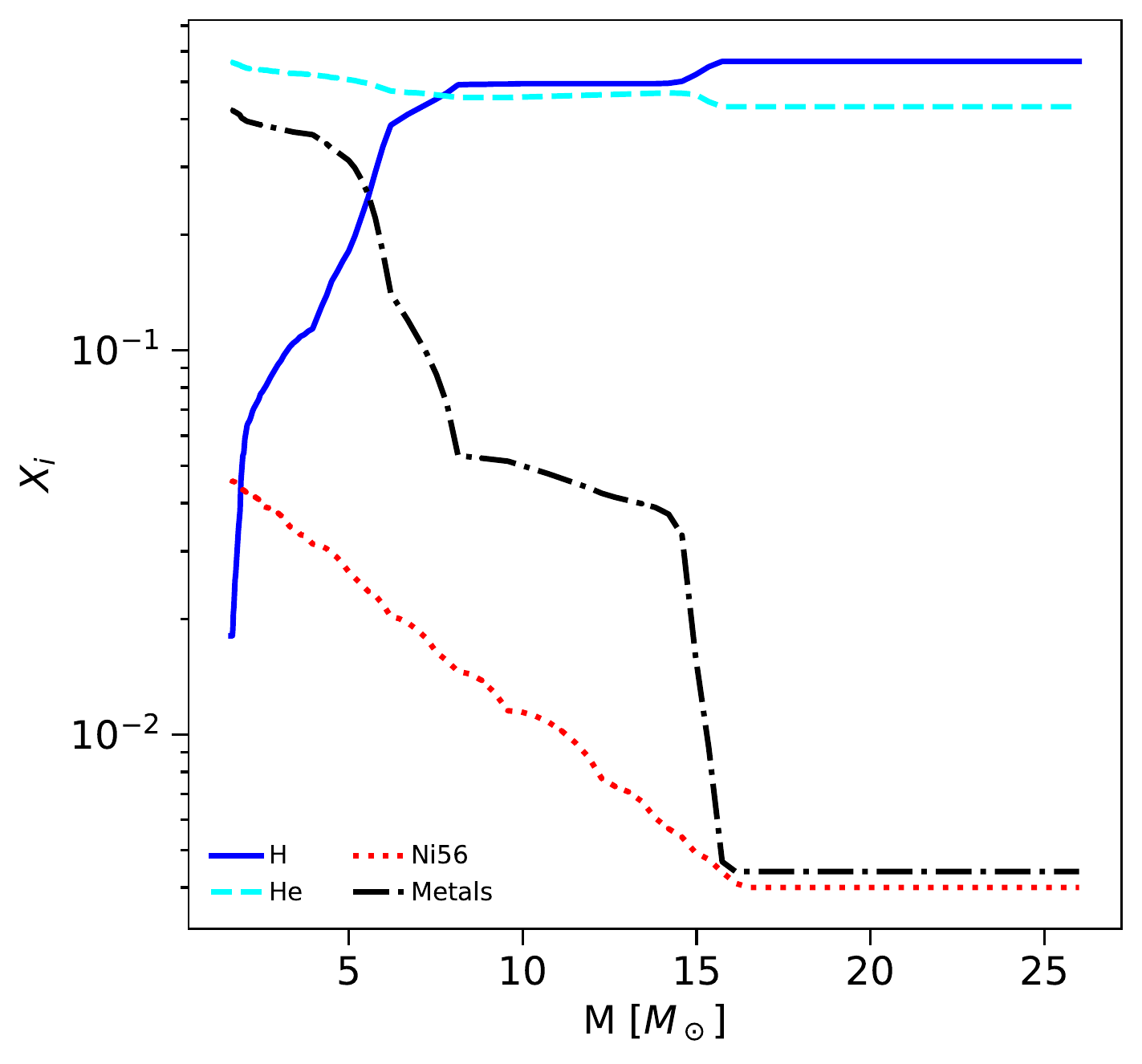}
 \includegraphics[width=0.5\textwidth]{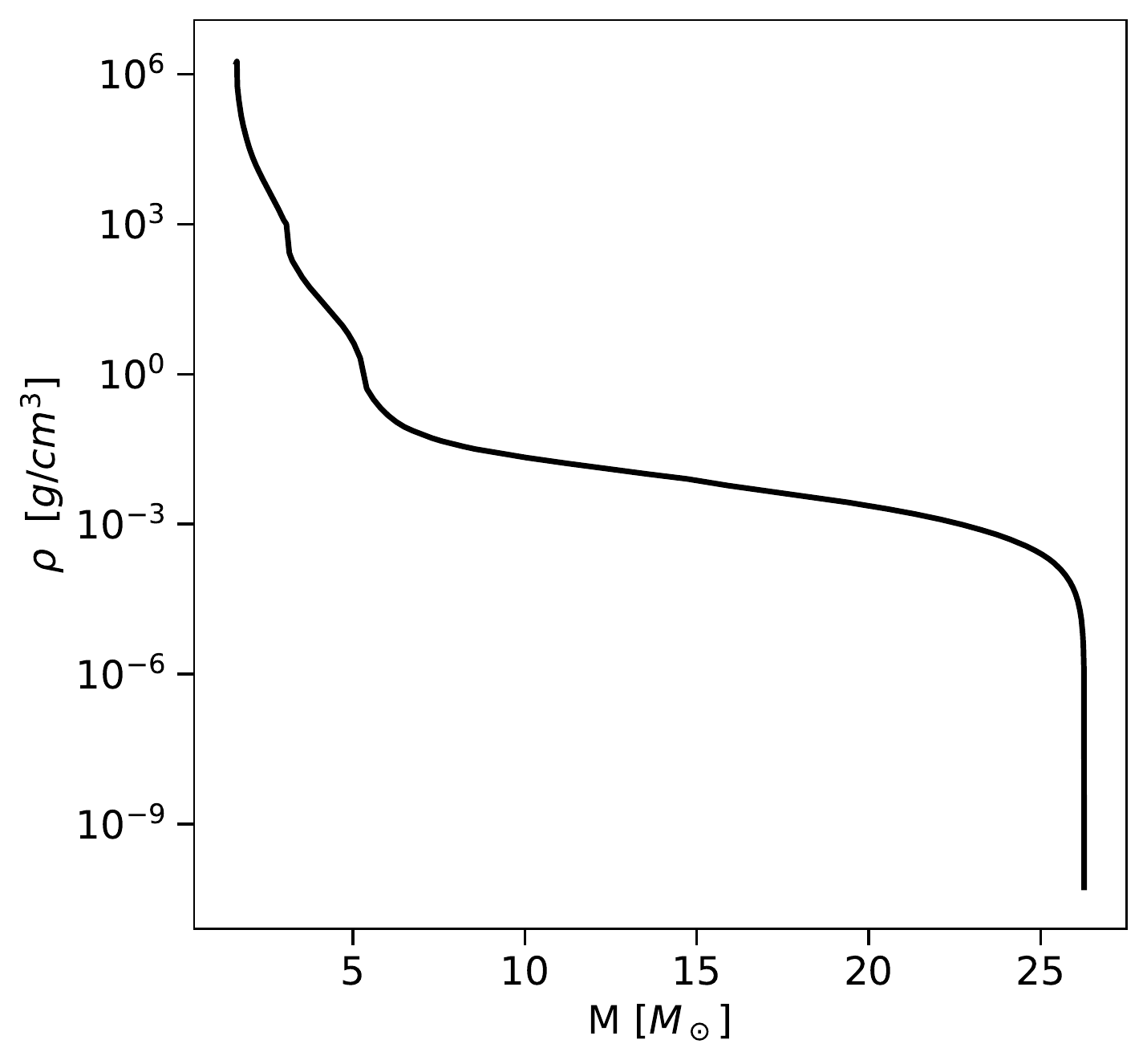}
}
	\caption{ Left panel: The mass fraction of hydrogen (blue solid line), helium (cyan dashed line), 
	heavy elements (black dash-dotted line), and radioactive \nifsx (red dotted line) in the ejecta of the model M1 
	(the model which best reproduces the observed SN Refsdal light curves; see Section~\ref{subsec:best_fit_model}).
	Right panel:  the density distribution of the model M1 as a function of interior mass.
	The total mass of the pre-supernova is $M_{tot} = 26.3\,M_{\odot}$, $M_{^{56}\mathrm{Ni}}=0.25\,M_{\odot}$, the  mixing is artificial.
	After an explosion, a \edit1{proto-neutron star}  core with a mass of $1.6\,M_{\odot}$ is left.}
\label{fig:chem_refR50M26}
\end{figure*}

As the initial model of chemical composition and density profile, we use the  well-studied
pre-supernova model of  \cite{Nomoto1988} \edit2{and \cite{Saio1988}, 
	and the explosive nucleosynthesis model of \cite{Shigeyama1990b}}. 
\cite{Blinnikov2000:sn1987a} performed a detailed analysis of SN 1987A and showed that an explosion of the evolutionary model of  \cite{Nomoto1988}  allows \edit1{one} to reproduce with enough precision \sna\ light curves and dynamical properties of expanding shell.
%
%Mixing of \nifsx and hydrogen is an important ingredient of a pre-SN model since it has a significant impact on the shape of a light curve \cite{Shigeyama1990b}.
Mixing of \nifsx\ \edit2{and hydrogen} is an important ingredient of a pre-SN model since it has a significant impact on the shape of a light curve \edit2{and on the observed gamma and X-ray radiation of SN1987A  \citep{Bartunov1987,Kumagai1989,Shigeyama1990b}}.
To obtain a light curve with a broad dome-shaped maximum, one needs to mix \nifsx closer to the edge of the envelope \edit2{ and hydrogen} down to the central region. 
Then the radioactive decay of \nifsx would start heating and ionizing a material at the shell edge just after the shock breakout.
\edit2{And the hydrogen-recombination front would exist for a longer period.}
This would cause an increase in the photosphere radius. 
Note that a similar approach to mixing of \nifsx was used in \cite{Utrobin2011a}  to explain the light curve and spectroscopic data of  SN\,2000cb which is also \edit1{a} peculiar  \sna-like SN  and   characterized by a wide dome-like light curve maximum.  
\edit1{Following \cite{Baklanov2005}, we constructed the blue supergiant (BSG) models in non-evolutionary hydrostatic equilibrium by  varying masses of hydrogen and helium in the outer shell. }
Typical chemical composition and density distribution in our SN models are shown in~Figure~\ref{fig:chem_refR50M26}.

The explosion was triggered using the ``thermal bomb'' model \citep{Shigeyama1990b,Blinnikov2000:sn1987a},
namely, via a short ($\Delta t_{burst}=0.1 s$) release of thermal  energy $E_{burst}$ in 
the near-central \edit1{layers} with the mass of $0.06 \msun$ on the outer edge of the core
with the mass of $M_{core} = 1.6 \msun$ (\cite{Blinnikov2000:sn1987a}). 
The core material  forms a proto-neutron star and does not participate in the expansion of the supernova envelope. 
In the equations of motion of the envelope material, the contribution of the core to the gravitational potential is taken into account.

\begin{figure*}[htb!]
    \centerline{
        \includegraphics[width=0.5\textwidth]{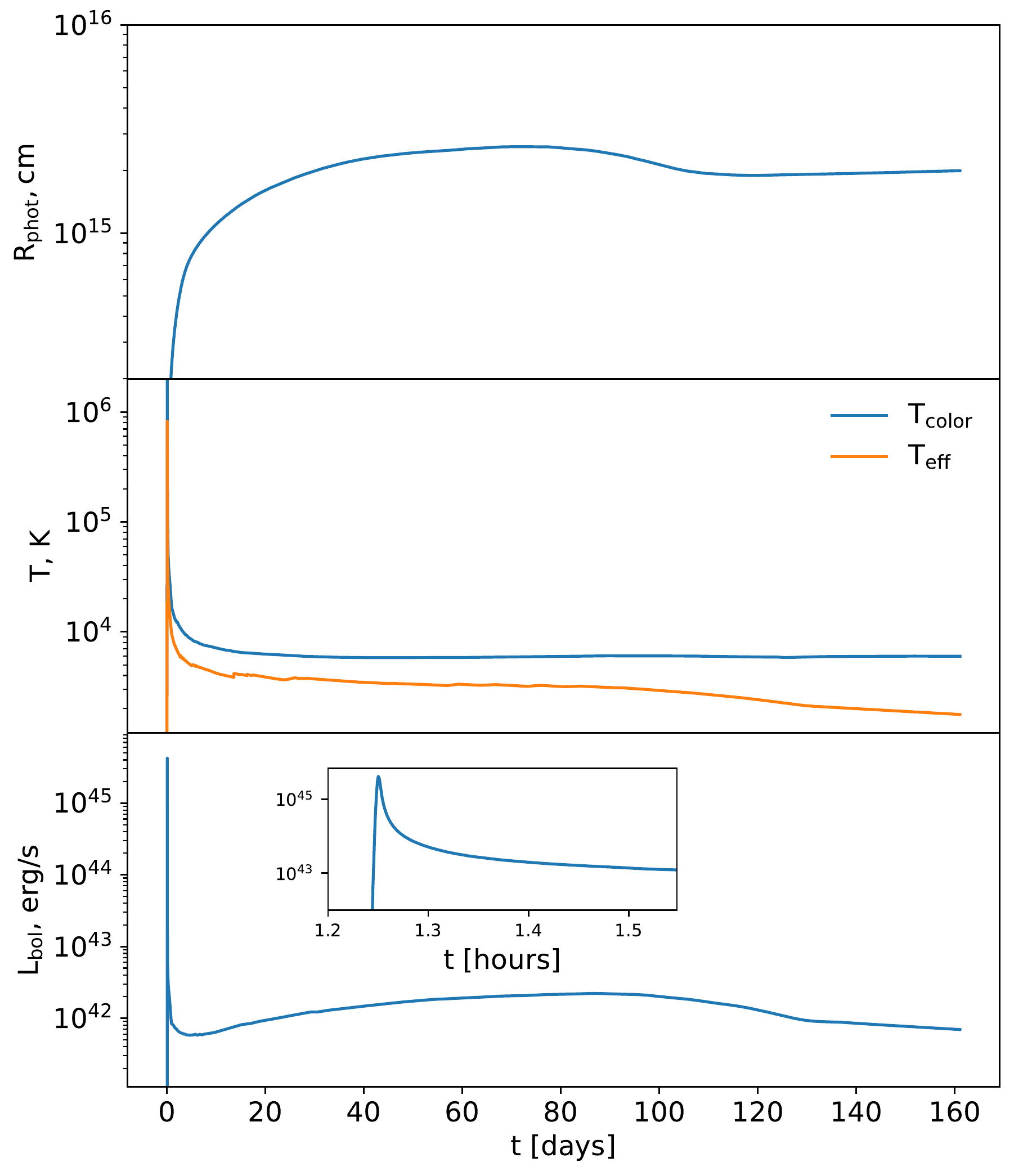}
        \includegraphics[width=0.5\textwidth]{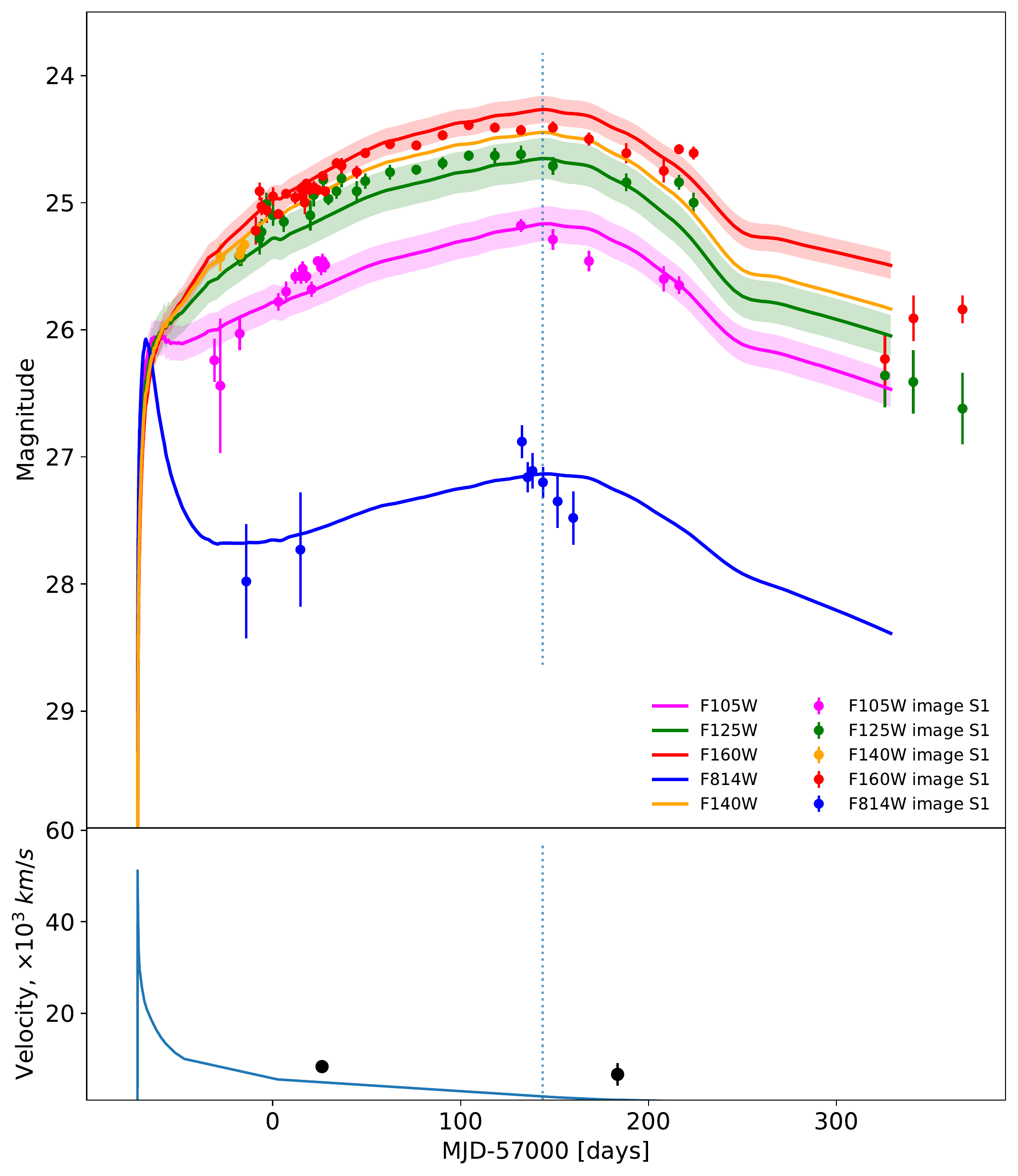}
    }
	\caption{The observational properties of the best-fit model M1.   
	The left panel shows the photospheric radius, the color and effective temperature, and the bolometric luminosity. 
	\edit1{The upper panel on the right} shows the multi-band photometry  for S1 image 
	and \edit1{the lower one demonstrates} the evolution of the photosphere velocity in
	comparison with  H$\alpha$ expansion velocity measurements (black circle) from \citealt{Kelly2016}. 
	\edit1{The vertical dotted line marks the date of the peak brightness in F160W.} 
	Note that observed LCs in all filters  are well-reproduced by M1, although for fitting 
	we used only F160W, F125W, and F105W passbands (these filters are shown with the model uncertainties).}
	\label{fig:RphTLbol_refE5R50M26Ni2m2b5m3Z01}
\end{figure*}

\renewcommand{\arraystretch}{1.2}
\begin{table*} 
\centering
	\caption{The parameter space of our 185 SN models}
	\begin{tabular}{cccccc}
		\hline \hline    & $R_0$ [\rsun] &   $M_{tot} $ [\msun] & $M_{^{56}\mathrm{Ni}}$ [\msun] & $E_{burst}$ [E51] & $Z$  \\ 
		\hline min & $ 30 $ & $ 16 $ & $0.077 $ & $ 1 $ &  \edit1{0.001} \\ 
		& \vdots  &  \vdots  &  \vdots  &  \vdots &  \vdots  \\ 
		max & $ 100$ & $ 27 $ & $0.42 $ & $ 7$ &   \edit1{0.005} \\ 
		\hline 
	\end{tabular}  \label{tbl:space_params}
\tablecomments{The parameter space of our 185 SN models: 
minimum and maximum values of the radius and total mass of \edit1{the} pre-supernova,
\nifsx mass, the total explosion energy (in units of $10^{51}$ erg), and the metallicity,
\edit1{defined as the mass fraction of elements heavier than helium in the outer shells of a pre-supernova}.}
\end{table*}

\begin{table*}%[h]  
\centering
	\caption{\edit2{The mean values of SN model parameters computed using the Bayesian Model Averaging method and} the parameters of the best-fit model M1. }
	\begin{tabular}{ccccccc}
		\hline\hline    & $R_0$ [\rsun] &  $M_{tot}$ [\msun]&  $M_{ej}$ [\msun] & $M_{^{56}\mathrm{Ni}}$ [\msun] & $E_{burst}$ [E51] & $Z$  \\ 
	 \hline \edit2{BMA}  & $ 50\pm 1 $ & $ 25\pm2 $ & $ 23.4\pm2$ & $0.26\pm0.05$ & $ 4.7\pm0.8$ &  $0.005\pm 0015$ \\ 
		 Best-fit &  $ 50 $ & $ 26.3 $ & $ 24.7 $ & $0.247 $ & $ 5 $ &  $ 0.004$ \\ 
		\hline 
\end{tabular} \label{tbl:mbest_params1}
%\tablecomments{The parameters  of the best-fit model M1: the radius and total mass of a pre-supernova, the ejecta mass, the radioactive \nifsx mass, the total explosion energy (in units of $10^{51}$ erg), and the metallicity.} 
\end{table*}

To obtain the model which simultaneously reproduces multi-band SN Refsdal light curves,
we computed \edit1{a}  set of  185 radiation-hydrodynamical models. 
SN Refsdal \edit2{is} \edit1{a} gravitationally lensed supernova, and the absolute magnifications of its images are poorly constrained. 
For example, the absolute magnification for S1 predicted by different lens 
models\footnote{ \edit1{Note that the  range of magnifications provided here 
does not appear to be exhaustive, since absolute magnifications are not available
in literature for all the lens models.}} \citep{Oguri2015, Kawamata2016, Sharon2015, Jauzac2015, Grillo2016} is  $\mu_{S1}\sim 10\div \edit1{25}$. 
Thus, we \edit1{cannot} use the observed peak luminosity as a constraint.
Instead, we try to reproduce the shape of SN Refsdal light curves in different bands 
keeping in mind    available in the literature estimates of the absolute magnifications 
for S1 and measured in \cite{Kelly2016} H$\alpha$ expansion velocity.
The ranges of values of the SN models parameters are given in  Table~\ref{tbl:space_params}. 
Note that the parameters  are not distributed uniformly in the parameter space, but converge to some optimal model (in the sense of Bayesian evidence, see  Appendix~\ref{adx:Bayesian}).
At each time step, \stella calculates the spectral energy distributions (SEDs) which are then transformed 
from host galaxy rest-frame ($ z = 1.49$)  into the  observer's frame  and
 convolved with the transmission functions of the HST filters. 
 Here, we use the F105W, F125W and F160W bands since
  the  coverage  of  the  Refsdal  light  curve in these bands is most complete and well-sampled.

\subsection{Best-fit SN Refsdal model} \label{subsec:best_fit_model}
\edit1{Our procedure to find}  the best-fit model (as well as  the time delays and magnification ratios) is described in detail in Appendix~\ref{adx:Bayesian}. 
Here, we just outline the main steps.
For each computed SN model, we compare synthetic LCs with observations and maximize the likelihood  function~\eqref{eq:likelihood}  with five free parameters: 
the absolute time and magnitude shifts for the reference image S1, and the model photometric uncertainties in three HST passbands used
(see Appendix~\ref{adx:Bayesian} for a discussion why model uncertainties are introduced as fit parameters). 
As the priors  for all the fit parameters, we use uniform distributions spanning a wide range of values. 
Then we determine the time and magnitude shifts of images S2-S4 relative to S1 in a similar fashion.
Next, we calculate the posterior probability of each SN model, and use the obtained value as a measure of how well the model fits observations. 
We report the 8 best-performing SN models  \edit2{with corresponding posterior probabilities} in Table~\ref{tbl:full_param_models} and show the model light curves in comparison with observations in Figure~\ref{fig:ubv_obs_models_S1S4}. 
\edit2{
To obtain the mean pre-SN parameters and estimate their uncertainties, we use the Bayesian Model Averaging approach 
(BMA; \citealt{Hoeting1999}), which basically provides a weighted average for each parameter
of interest, incorporating the posterior probabilities in the weighting
(see Appendix~\ref{adx:Bayesian} for details). 
The BMA results are provided in Table~\ref{tbl:mbest_params1}. 
Since the model M1   significantly outperforms all other explored models (see column 2 of Table~\ref{tbl:full_param_models}), the BMA values are quite close to the best-fit parameters.  
%, since the model M1 has the highest  posterior probability among the set of explored models (see Table~\ref{tbl:full_param_models}).
Below we discuss in detail particular models constructed in this work.  
}
% As follows from values in column 2 of Table~\ref{tbl:full_param_models}, the model M1  significantly outperforms all other explored models.

The model M1  fits best to the observed SN Refsdal LCs, and the resulting photospheric velocity is in agreement with available H$\alpha$ expansion velocity measurements from \cite{Kelly2016}.
The photospheric velocity can be inferred from  a blueshift of weak absorption lines such as  the lines of FeII~5018\AA\ and 5169\AA. 
For SNe type IIP including peculiar 1987A-like SNe, the FeII  lines show systematically lower velocities compared to H$\alpha$ \citep{Blanco1987, Taddia2012}.
Therefore, H$\alpha$ velocities should be systematically higher than the photospheric velocities of our models.
The main parameters of the best-fit model M1 are the following:
the total mass $M_{ej} = 26.3 \,M_{\odot}$,
the ejecta mass $M_{ej} = 24.7 \,M_{\odot}$,
the pre-SN  radius $R_0=50 \, R_{\odot}$,
the \nifsx mass $M_{^{56}\mathrm{Ni}} = 0.25 \,M_{\odot}$
and the explosion energy $E_{burst} = 5\times10^{51}$\,ergs (listed  in Table~\ref{tbl:mbest_params1}).
%

% \edit2{To estimate uncertainties on the pre-SN parameters, we exploit information on the whole set of constructed SN models. We compute the  mean and uncertainty for each parameter
%  (see Table~\ref{tbl:mbest_params1}) 
% using the Bayesian Model Averaging approach (BMA; \citealt{Hoeting1999}), 
% which basically provides a weighted average for each parameter of interest,
% incorporating the posterior probabilities in the weighting 
% (for details, see Appendix \ref{adx:Bayesian}, equations~\ref{eq:bma1}-\ref{eq:bma2}). As expected, BMA mean values are quite close to the best-fit parameters, since the model M1 has the highest  posterior probability among the set of explored models (see Table~\ref{tbl:full_param_models}). Below we concentrate not on the BMA result but on particular models constructed in this work. }

%\edit2{Based on the whole set of explored SN models, we also compute the average
%pre-SN parameters with uncertainties (see Table~\ref{tbl:mbest_params1}) 
%using the Bayesian Model Averaging approach (BMA; \citealt{Hoeting1999}), 
%which basically provides a weighted average for each parameter of interest,
%incorporating the posterior probabilities in the weighting 
%(for details, see Appendix \ref{adx:Bayesian}, equations~\ref{eq:bma1}-\ref{eq:bma2}). 
%The average values are quite close to the best-fit parameters, since the model M1 has the highest  posterior probability among the set of explored models.
%The parameters are listed  in Table~\ref{tbl:mbest_params1}
%}

BSGs are compact, and  the time of shock breakout is $\approx$\,1.25 hour  with  the boundary velocity of $\approx$\,59\,000 km\,s$^{-1}$ (see Figure~\ref{fig:RphTLbol_refE5R50M26Ni2m2b5m3Z01}).
Soon after the shock breakout at $t\approx$\,1.4 hour the radiative losses became small compared to the kinetic energy of the shell. 
Thus, internal temperature in the shell falls almost adiabatically, while the  bolometric luminosity decreases to $7\times10^{41}$ ergs/s and then reaches its local minimum at day 1 after the explosion (see Figure~\ref{fig:RphTLbol_refE5R50M26Ni2m2b5m3Z01}, left panel).
Details of the explosion model (the duration of the energy release and the mass of 
the thermal bomb)  majorly influence the magnitude and the shape of the first maximum of 
the light curve.
For the \snref modeling, these parameters are relatively inessential, 
 since   observations started during the  rise toward the second maximum of the light curve (cupola), 
 which forms determined by properties of the cooling and recombination wave and contribution of radioactive  \nifsx decay. 
 
Our  $M_{^{56}\mathrm{Ni}} = 0.25 \,M_{\odot}$  estimate is within the range of $(0.005 - 0.28)M_{\odot}$ implied by
 the observed distribution of $M_{^{56}\mathrm{Ni}}$ for  SN\,II \citep{Muller2017}
and \edit1{is} consistent with the high  energy of explosion \citep{Sukhbold2016}.
Note that the total energy release in M1 is greater than $ 2\times10^{51}$ erg, i.e. beyond the range  implied by neutrino-driven explosions \citep{Sukhbold2016}.
\cite{Sukhbold2016}   found  that at  most  6-8\%  of  the  SN\,IIP  explosions  are connected  to progenitors more massive than 20\msun when they used SN 1987A-calibrated neutrino engines. 
\edit1{ 
	However, a typical evolutionary scenario for a single star  with the total mass of $\sim26$\msun\ does not lead to a BSG stage before a supernova explosion.
	The low metallicity models	\citep{Hillebrandt1987} or 
	a merger of a compact binary system  \citep{Menon2019} could explain the appearance of the BSG pre-supernova.% models before their explosion. 
}
% The best-fit model M1 has the ejecta mass and the burst energy that are difficult to reconcile with a single star evolution scenario.
%
Nevertheless, M1 with $M_{ej} = 24.7\, M_{\odot} $ is quite similar to other  well-explored peculiar 1987a-like SNe, \edit1{with many of them having}  $M_{ej}> 20\, M_{\odot}$ \citep{Taddia2012}.

Metallicity\footnote{\edit1{Throughout this work, metallicity $Z$ is defined 
%as the mass fraction of metals in the outer layers of a star}
as the mass fraction of elements heavier than helium in the outer shells of a progenitor}} in the outer layers of the M1 envelope is reduced by the factor of \edit1{5} ($Z_{M1} = \edit1{0.004}$) relative to \edit1{ the solar metallicity }($Z_\odot = 0.02$) and is comparable with the value  ($Z = 0.005$) of the model by \cite{Nomoto1988}. 
\edit1{The decline in the metallicity in the outer layers leads to the reduction of the line opacity \edit1{which is calculated }  in \stella as  expansion opacity \edit1{following} \cite{Eastman1993}.}
%The low metallicity decreases the line opacity, \edit1{which is calculated }  in \stella as  expansion opacity \edit1{following} \cite{Eastman1993}.
\edit1{While the cooling and recombination waves are propagating},  the contribution of line opacity to the total opacity is significant.
The line opacity drops dramatically from UV toward optical wavelengths. 
Thus, lower metallicity mainly affects the light curves  in blue and UV light curves, what allowed us 
 to reproduce the light curves in the F814W band (see Figure~\ref{fig:RphTLbol_refE5R50M26Ni2m2b5m3Z01}, right panel).
%
% The low metallicity \snref  may be explained by the formation of a star in the early universe.

According to our best-fit model, the absolute magnification of S1 image is  $\simeq 10$
(see column~8 of Table~\ref{tbl:models_dt_dm}) 
which \edit2{is within}  the range of magnifications $\mu_{S1}\sim10\div \textbf{25}$
predicted by different lens models \citep{Oguri2015, Kawamata2016, Sharon2015, Jauzac2015, Grillo2016}. 
To increase the absolute magnification of S1, we should reduce a radiated flux. 
The model M4 (see Table~\ref{tbl:full_param_models}) has $M_{^{56}\mathrm{Ni}} = 0.12\, M_{\odot}$ in the envelope, i.e. $\sim2$ times less than the best-fit model. 
Thus, the amount of heat released as a result of  \nifsx decay is also  $\sim2$ times lower,
and smaller amount of energy can be radiated. 
Figure~\ref{fig:ubv_obs_models_S1S4} illustrates that M4 fits the SN Refsdal LCs if the absolute magnification of S1 is $\mu_{M4} = 18.6$, i.e. $\sim2$ times larger than for M1. 
In principle, the amount of \nifsx and, as a consequence, the absolute magnification, can be constrained from spectral lines  \citep{Utrobin2011a}  or by
the  slope of the tail of the supernova light curves \citep{Nadyozhin2003}.

% % % % % % % % % % % % % % % % % % % % % % % % % % % % % %
\section{Time delays and magnification ratios}
\label{sec:dt_and_mu}
\subsection{SN Refsdal images S1-S4 }

In the previous Section, for each computed SN model we derived the best-fit absolute time and magnitude shifts  for the image S1. 
By fitting the model  light  curves to images S2, S3, and S4, we obtained the time shifts and magnifications of images S2-S4 relative to  S1 (for details, see Appendix~\ref{adx:Bayesian}).  
Figure~\ref{fig:ubv_fitbands_refE5R50M26Ni2m2b5m3Z01} illustrates the results of our time delay and magnification calculations  for the best-fit model M1. Each panel shows the composite light curve from images S1–S4, after applying the time and magnitude shifts so that S2-S4 match the S1 light curve. 
The best-fit model light curves are overplotted as red (F160W filter), green (F125W) and magenta (F105W) lines with the  shaded bands indicating the  model uncertainty. 
The photometric model uncertainties in each filter for the best-fit model are provided in Appendix~\ref{adx:Bayesian}, Table~\ref{tbl:full_param_models} (see the first row).

\begin{table*}[htb] 
	\centering
	\caption{Summary of time delay and magnification ratio measurements.} 
	\begin{tabular}{cccc}
		\hline\hline  
		Parameter &                 BMA Mean       &  Template Fits\tablenotemark{a} & \edit1{Polynomial}  Fits\tablenotemark{a}  \\ 
		 $\mathtt{MJD}_{pk}$  & $ 57144^{+10}_{-10} $\,d & $ 57138\pm10$ d & $ 57132\pm4 $ d \\ 
		 $\Delta t_{S2-S1}$   & $ 9.5^{+2.6}_{-2.7}  $ d & $ 4\pm4 $ d  & $7\pm2 $ d \\ 
		$\Delta t_{S3-S1}$   & $  4.2^{+2.3}_{-2.3} $ d & $ 2\pm5 $ d  & $ 0.6\pm3 $ d \\ 
		$\Delta t_{S4-S1}$   & $ 30^{+7.8}_{-8.2} $ d & $ 24\pm7 $ d & $ 27\pm8 $ d \\ 
		$\Delta t_{SX-S1}$   & $ 340^{+43}_{-52} $ d & --- & --- \\ 
		\hline $\mu_{S2/S1}$        & $ 1.14^{+0.021}_{-0.020} $  & $ 1.15\pm0.05 $ & $ 1.17\pm0.02 $ \\ 
		$\mu_{S3/S1}$        & $ 1.01^{+0.019}_{-0.018} $  & $ 1.01\pm0.04 $ & $ 1.00\pm0.01 $ \\ 
		$\mu_{S4/S1}$        & $ 0.35^{+0.016}_{-0.015} $  & $ 0.34\pm0.02 $ & $ 0.38\pm0.02 $ \\ 
		$\mu_{SX/S1}$        & $ 0.24^{+0.12}_{-0.07} $  & --- & --- \\ 
		\hline 
	\end{tabular}  \label{tbl:dtmu}
	
	\tablenotetext{a}{ values from \citealt{Rodney2016}, see their Table 3.}
	\tablecomments{Summary of time delay and magnification ratio measurements. The second column presents results obtained in this work in comparison with estimates obtained  in \citealt{Rodney2016} via light curve template fitting and polynomial fits.} 
\end{table*}

\begin{figure*}[htb!]
  \centering
	\includegraphics[width=\linewidth]{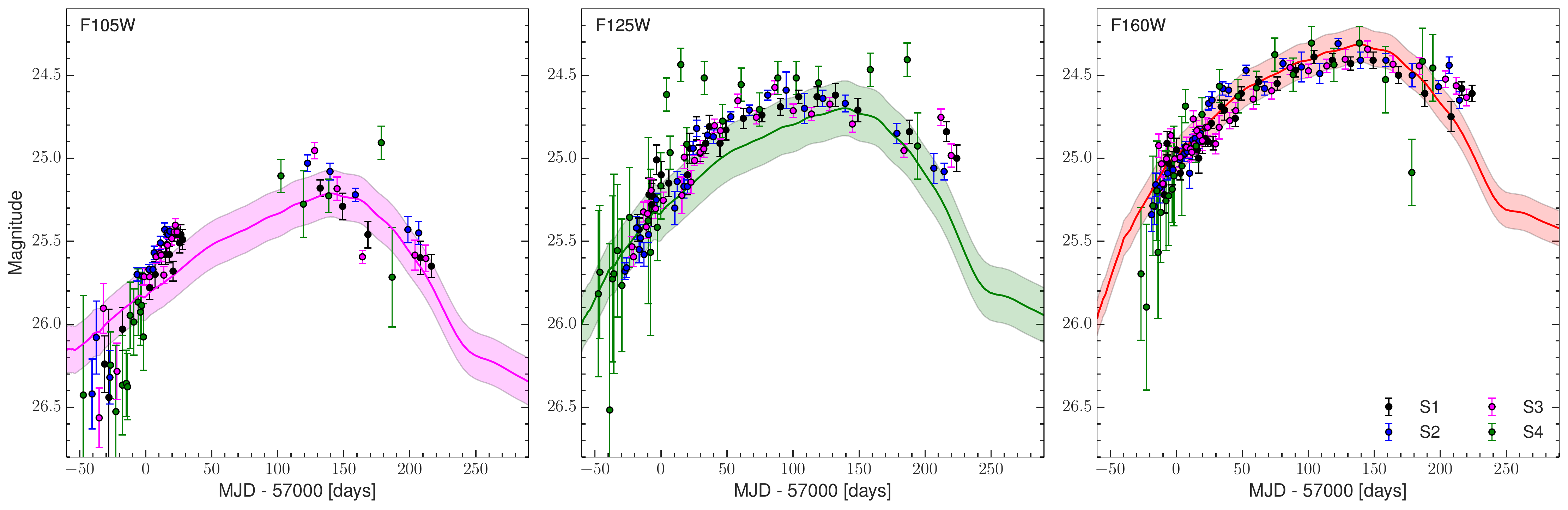}
	\caption{The composite light curves from  images S1-S4 after applying the magnitude and time shifts (relative to the image S1) determined for the best-fit model M1.  \edit1{Measurements for S1-S4 are shown, respectively, in black, blue, magenta, and green.} The best-fit model light curves are shown as solid lines with photometric uncertainties as shaded areas. } 
	\label{fig:ubv_fitbands_refE5R50M26Ni2m2b5m3Z01}
\end{figure*}

To derive a single measurement of the time delay and magnification ratio for SN images, that takes into account the Bayes factor and the uncertainty of each SN model,  we use the \edit2{BMA method} (equations~\ref{eq:bma1}-\ref{eq:bma2}).
%Bayesian Model Averaging approach (BMA; \citealt{Hoeting1999}), which basically provides a weighted average for each parameter of interest, incorporating the posterior probabilities in the weighting (equations~\ref{eq:bma1}-\ref{eq:bma2}). 
Mean values of time delays and magnification ratios obtained from the BMA combinations are provided in Table~\ref{tbl:dtmu}.

\begin{figure*}[ht] 
	\centering
	\plotone{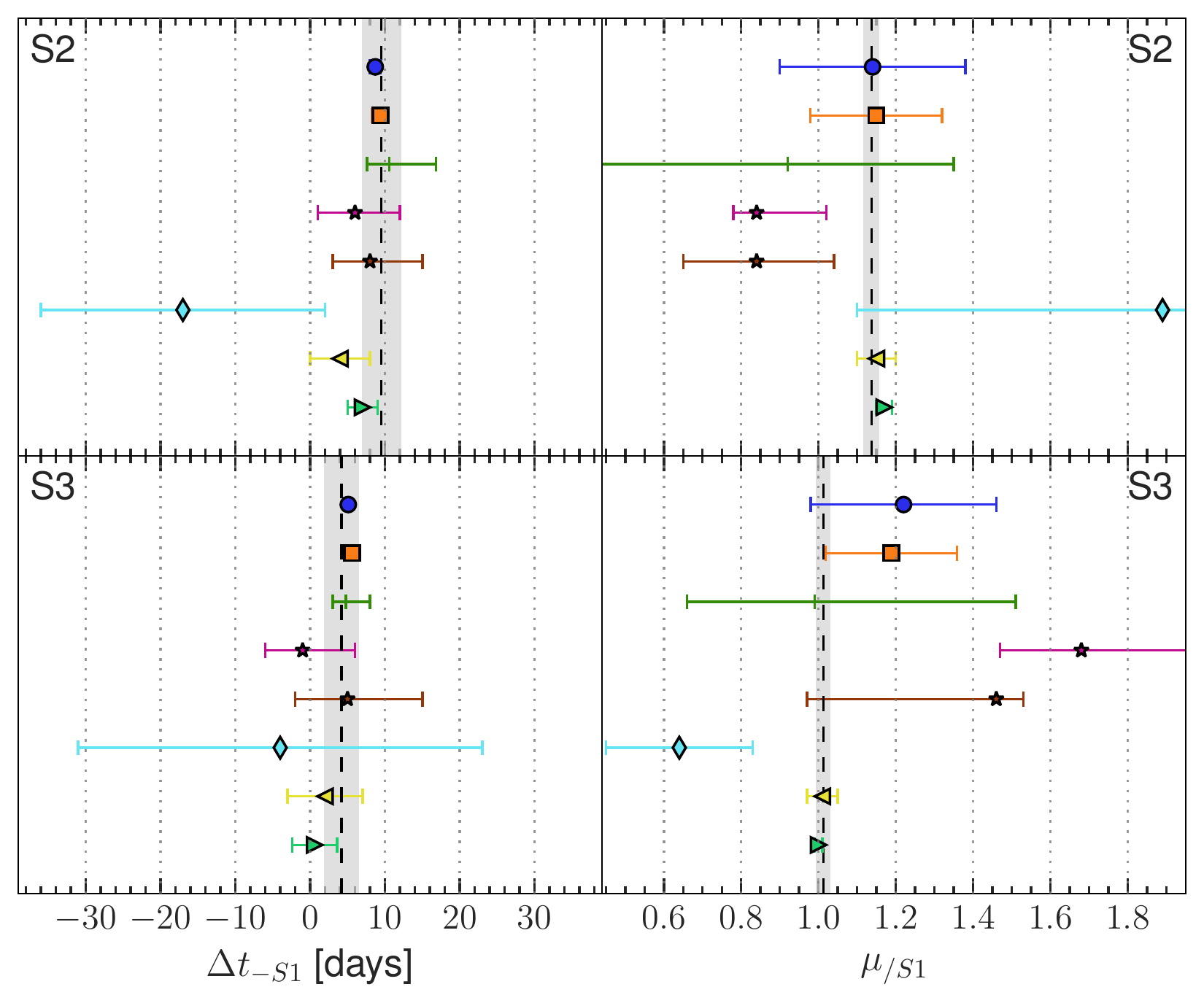}
	\plotone{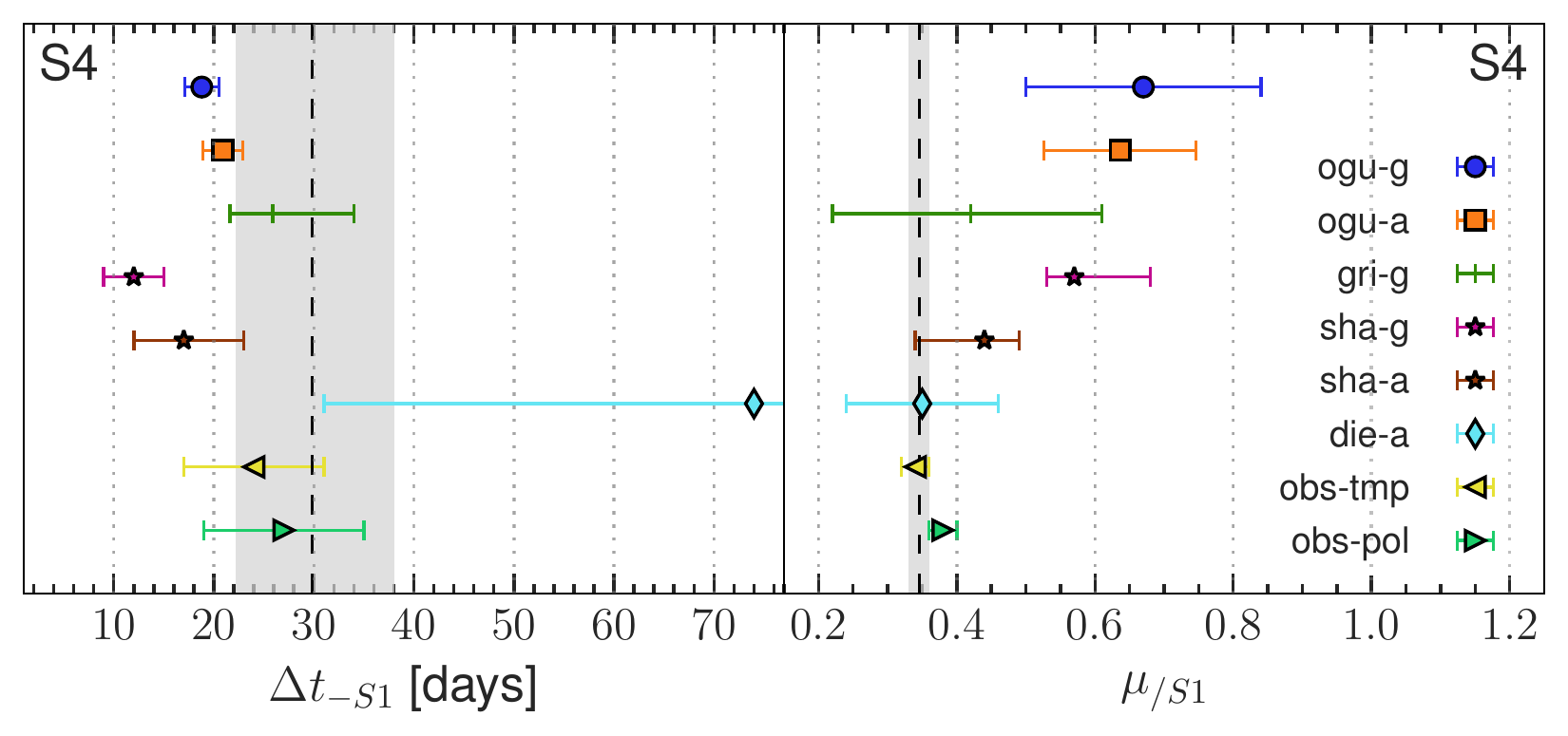}
	\caption{ Comparison of lens model predictions \edit1{from the literature} with time delay and magnification values obtained in this work. 
		The left and right columns present the time delays and magnification ratios (both relative to S1), correspondingly. Results for images S2, S3, and S4 are shown from top to bottom. The dashed vertical lines and the gray shaded regions indicate  our best estimate of the time delay/magnification ratio with uncertainties (see  column 2 in Table~\ref{tbl:dtmu}). Values predicted by different lens models (`Die-a',`Gri-g', `Ogu-a', `Ogu-g', `Sha-a', `Sha-g') are plotted as colored \edit1{symbols} (see the legend  in the lower right panel). Yellow and green triangles mark the measurements from \citealt{Rodney2016} derived from light curve template (`obs-tmp') and polynomial (`obs-pol') fitting, correspondingly.  }
		\label{fig:gl_dtdm_S2S4}
\end{figure*}

Thanks to the discovery of the first multiply-lensed supernova, the galaxy cluster  MACS~1149.5+2223 has been extensively observed and modelled by several independent lens teams \citep[see, e.g.,][]{Oguri2015, Kawamata2016, Sharon.Johnson2015, Grillo2016, Jauzac2015}. Comparison of lens models and summary of the time delays and magnification ratios predicted by those models are given in \cite{Treu2015}. 
Figure~\ref{fig:gl_dtdm_S2S4} presents a comparison of our measured mean time delays and magnification ratios for SN Refsdal images S1–S4 against the lens model predictions from \cite{Treu2015} (namely, `Die-a',`Gri-g', `Ogu-a', `Ogu-g', `Sha-a', `Sha-g'). \edit1{The estimates of the time delay and magnification ratios resulting} from the template and polynomial fitting \citep{Rodney2016} are also shown. Our \edit1{time delays and magnification ratios of the SN Refsdal multiple images S1-S4 are consistent, within the errors,}  with results of  \cite{Rodney2016} (see  Table~\ref{tbl:dtmu}).

\subsection{SX} \label{subsec:SX}

Approximately a year after the discovery of SN Refsdal `Einstein cross', a fifth image appeared. As it was expected, SX is much fainter than S1-S4 and its photometric measurements are scarce. 
\edit1{To place constraints on the  time delay and magnification of SX relative to S1,
we, on the one hand, use available SX photometry from \cite{Kelly2016_SNX} (their Table~1) in F125W and F160W filters and, on the other one, information about  SX brightening,  registered in January 2016 \citep{Kelly2016_SNX}. }
We repeat the procedure of maximizing the likelihood function (see Appendix~\ref{adx:Bayesian}) to find the best-fit values of the time and magnitude shifts of SX relative to S1 for  8 best-fit SN models. 
The 8 best performing models are illustrated in Figure~\ref{fig:ubv_SX_fit_models_t0}, and Figure~\ref{fig:triangle_all} shows  the marginal distributions for the SX-S1 time delay and the SX magnification for the best-fit model M1. 
The obtained marginalized distributions are quite broad (for all best performing models, not only for M1), and, as a consequence, uncertainties on the parameters of interest are large.
Due to a broad peak the model light curves  in F125W and F160w filters are relatively featureless, and 2-3 data points are not enough to obtain tight constraints.   

\begin{figure*}
\includegraphics[width=\linewidth]{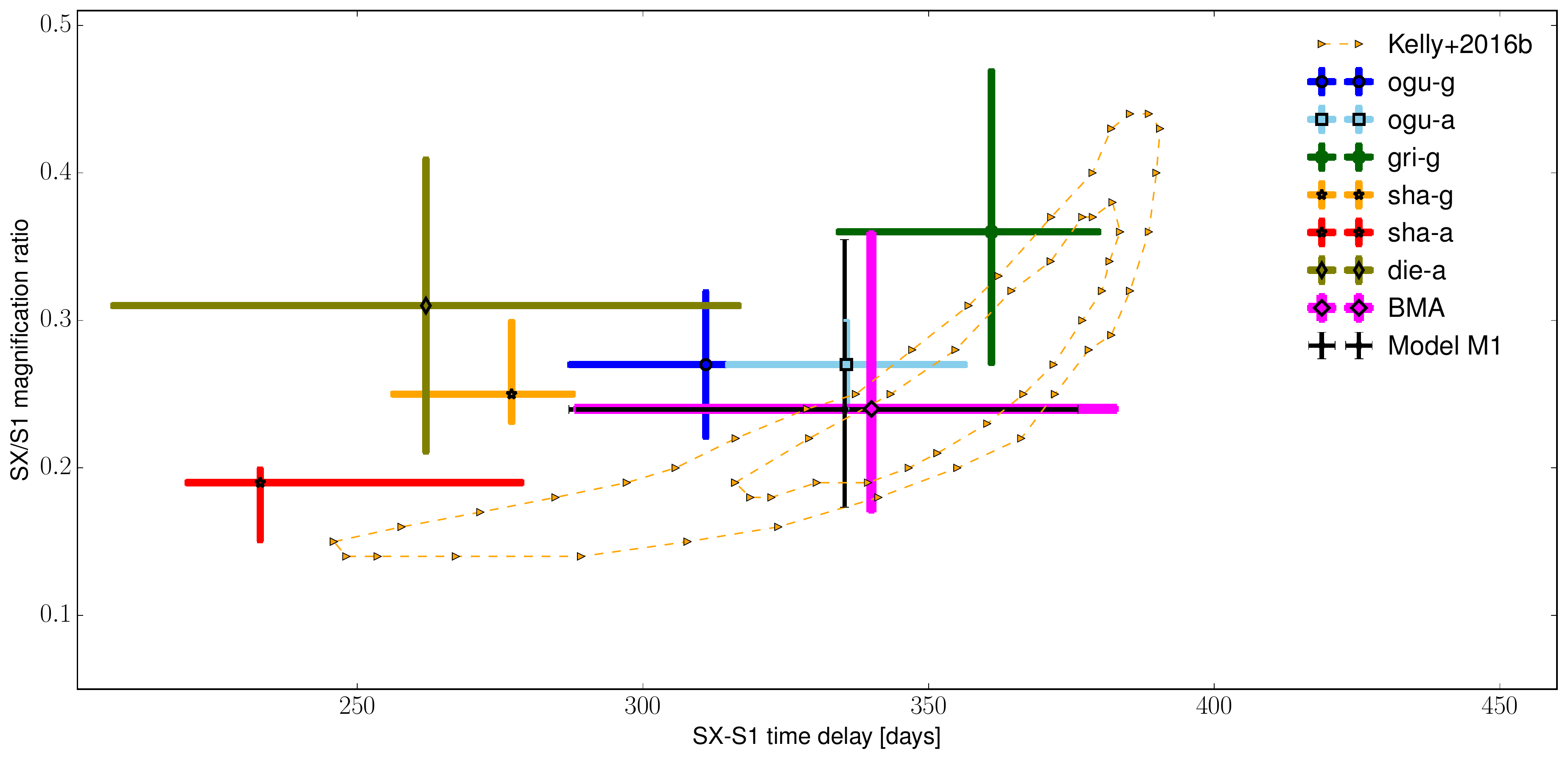}
	\caption{ The best-fit model M1 and  BMA constraints (in magenta) on the time delay and magnification of the image SX relative to S1. 1$\sigma$ and 2$\sigma$ orange contours show results from \citealt{Kelly2016_SNX} where the shape of LC was obtained via fitting second-order polynomials to S1 measurements in F125W and F160W bands separately. Values predicted by different lens models  are plotted as colored points (see the legend  in the right corner). } 
	\label{fig:SX_mcmc_dt_mu_lens_models}
\end{figure*}

We again calculate the mean values using the BMA method. 
The time delay and the magnification of SX are $\Delta t_{SX-S1}=  340^{+43}_{-52}$ days, $\mu_{SX/S1}=0.24^{+0.12}_{-0.07} $.
In Figure~\ref{fig:SX_mcmc_dt_mu_lens_models}, we plot the best-fit model M1  and BMA estimates of the time delay and magnification ratio between images S1 and SX in comparison with the constraints from \cite{Kelly2016_SNX} and lens model predictions  from several teams reported by \cite{Treu2015}.

\section{Hubble constant} \label{sec:H0}

More than half a century ago \cite{Refsdal1964} proposed to use time-delays between multiple images of gravitationally lensed supernovae to measure the Hubble constant. 
However, no multiply imaged SN has
ever been observed until just recently. In practice, the strong lens time delay cosmography has been employed extremely successfully for decades using multiply imaged quasars. 
For example,  the H0LiCOW collaboration \citep{Suyu2017} has recently constrained $H_0$ to 2.4\% precision for a flat $\Lambda$CDM cosmology from a joint analysis of six gravitationally lensed quasars with measured time delays \citep{Wong2020}. 
To achieve such a precision, one needs a  variety  of  observational  data. For example, to measure time delays between images several years of photometric  monitoring of  the  lens  system are typically required, because the
light curves of quasars are stochastic and heterogeneous and their intrinsic stochasticity is hard to disentangle from variability due to microlensing \citep[e.g.,][among others]{2005A&A...436...25E,2013A&A...556A..22T, 2015ApJ...799..168D}.
In contrast to quasars, gravitationally lensed SNe with multiple images occur on short timescales, allowing their time delays to be measured with far less observational efforts.  Moreover,  after  lensed SNe fade away, one can obtain imaging of a host galaxy to validate the lens model.
In addition, the intrinsic luminosities of SN Ia and \edit1{of} certain types of core-collapse SNe can be determined independently of lensing, which allows to directly measure the lensing magnification factor. 
A model-independent estimate of the magnification can improve constraints on the lens model especially for galaxy clusters with only few known multiple image systems \citep{Riehm2011}.

Here, we constrain the Hubble constant using the values of time delays between SN Refsdal images determined in Section~\ref{sec:dt_and_mu}. While modelling the SN Refsdal light curves and determining time delays between images, we have ignored the microlensing effect.
A preliminary assessment of whether there are any
indications of especially strong microlensing events that could
bias time delay and magnification measurements is given in \cite{Rodney2016}. They concluded that the SN Refsdal light curves are unlikely to be affected by major microlensing events. Throughout the paper, we assume that microlensing has no influence on our results but there are studies which show that microlensing does indeed introduce uncertainty in the time delay and the Hubble constant measurements \citep[see, e.g., ][]{Dobler2006,Goldstein2018,Pierel2019}. \cite{Huber2019, HOLISMOKES_I} discuss the best strategies to detect gravitationally lensed SNe and to measure their time delays with high accuracy in the presence of microlensing. 
\begin{figure*}
\includegraphics[width=\linewidth]{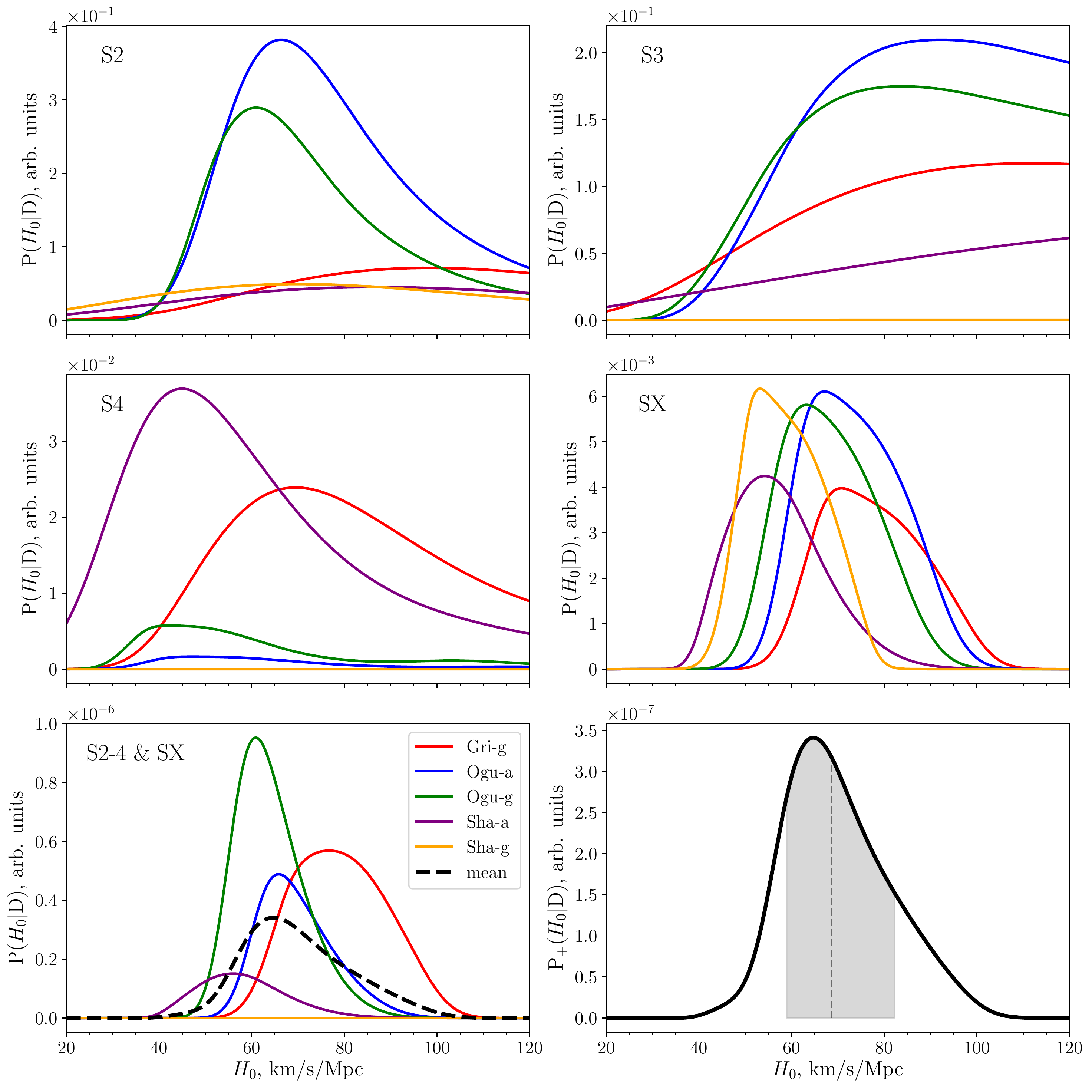}
\caption{$P(D|H_0) = \int p_{lens}(\Delta t | H_0) \cdot   p_{obs}(\Delta t)d\Delta t\, \times  \int 
p_{lens}(\mu)   \cdot   p_{obs}(\mu)d\mu\,$ for different lens models calculated for images S2, S3, S4, and SX separately.  Lens models are shown with different colors as indicated in the legend in the lower left panel. 
Lower left panel: \edit1{combined} contributions (from all images) of lens model predictions to the posterior distribution $\mathrm{P}_{+} (H_0 | \mathrm{D})$, shown in the lower right panel.  
Lower right panel: The total posterior distribution $\mathrm{P}_{+} (H_0 | \mathrm{D})$ defined as the mean of likelihood functions $P(D|H_0)$ of different lens models. 
The dashed line marks the median value of $H_0$. The grey shaded area denotes the 16th and 84th percentile confidence band. The best value for $H_0$ resulting from the combined analysis of all images  is $68.6^{+13.6}_{-9.7}$ km s$^{-1}$ Mpc$^{-1}$. } 
\label{fig:p_lens}
\end{figure*}

To derive $H_0$, we follow the approach  proposed in \cite{Vega_Ferrero2018} where the Hubble constant is obtained via re-scaling the time delays predictions of the lens models to match the observed values (although, see \citealt{Grillo2018} for discussion on possible caveats of this approach). The following lens models are considered here: `Gri-g', `Ogu-g', 'Ogu-a', 'Sha-g', and 'Sha-a'. Description of these models, time delays, and magnification predictions for all SN Refsdal images are given in \citealt{Treu2015}  (see their Table~6). 
Following \cite{Vega_Ferrero2018}, we estimate the probability of $H_0$ given `observational' data $\mathrm{D}$ (\edit1{i.e. the} obtained best-fit values of $\Delta t$ and $\mu$ listed in Table~\ref{tbl:dtmu}) as

\begin{equation}
\begin{split}
    \mathrm{P} (H_0 | \mathrm{D})  \propto \mathrm{P}(H_0) ~ \mathrm{P} (\mathrm{D} | H_0) \propto \\   
     \mathrm{P}(H_0) \int d\Delta t\, d\mu\,
  \cdot  p_{lens}(\Delta t,\mu | H_0)   \cdot   p_{obs}(\Delta t,\mu),
   \label{eq:post_models}
\end{split}   
\end{equation}

where $\mathrm{P} (H_0)$ is the prior for $H_0$ (assumed to be flat between 20 and 120 km s$^{-1}$ Mpc$^{-1}$), $p_{lens}(\Delta t,\mu)$ is the distribution of time delay and magnification predictions of a given lens model (which can be re-scaled to any alternative value of $H_0$), and $p_{obs}(\Delta t,\mu)$ is the `observational' distribution  obtained in this work. 
We assume that for each lens model, $p_{lens}(\Delta t,\mu$) is described with  a normal bivariate distribution (with no correlation between $\Delta t$ and $\mu$). 
The mean values and their statistical uncertainties for each image are taken from Table~6 of \cite{Treu2015}. 
To obtain the `observed'  distributions $p_{obs}(\Delta t)$ and $p_{obs}(\mu)$, we average marginalized distributions of all explored SN models using the BMA method with weights corresponding to  posterior probabilities of  SN models (see Appendix~\ref{adx:Bayesian}). 
Figure~\ref{fig:p_lens} shows the obtained $P(\mathrm{D} |H_0)$ for different lens models  for each image separately (upper and middle panels) and the `combined' $P(\mathrm{D} |H_0)$ distributions (lower left panel) calculated as a product of $P(\mathrm{D}|H_0)$ for separate images. 
The total posterior distribution $\mathrm{P}_{+} (H_0 | \mathrm{D})$ (see the lower right panel of Figure~\ref{fig:p_lens}) is calculated as the mean of `combined' $P(\mathrm{D}|H_0)$ distributions for lens models.  
The median value and 68\% for the Hubble constant are $68.6^{+13.6}_{-9.7}$ km s$^{-1}$ Mpc$^{-1}$.
We see that the lens models `Ogu-g', `Ogu-a', and `Gri-g' contribute most to the total $\mathrm{P}_{+} (H_0 | \mathrm{D})$, i.e. these models are in a good agreement with time delays and magnification ratios obtained in this work.

With more photometric measurements of \edit1{the} image SX , we will be able to drastically improve our time delay and magnification measurements as well as accuracy of $H_0$ determination.

% % % % % % % % % % % % % % % % % % % % % % % % % % % % % % % % % %
\section{Conclusions}
\label{sec:conclusions}

Hydrodynamic simulation of the light curves and the expansion velocities \edit1{allow} us to get significant insights into the nature of core-collapse SN progenitors, namely, to estimate \edit1{the} radius and \edit1{the} mass of a progenitor star, an explosion energy, an ejecta mass, and a radioactive \nifsx amount. 
At high redshifts, core-collapse supernovae, especially Type IIP, are hard to discover due to their faintness. 
The highest-redshift spectroscopically confirmed SN IIP is 
PS1-13bni  with  \edit1{a}  redshift  of $z=0.335^{+0.009}_{-0.012}$ \citep{Gall2018}. 
With the help of gravitational lensing, we can probe SN IIP at much greater distances. Before \edit1{the} discovery of SN Refsdal, the highest redshift core-collapse SN (most likely Type IIP) was at $z\simeq 0.6$ \citep{Stanishev2009}. 
This transient was found in the Abell 1689 galaxy cluster \edit1{field} and probably was magnified by $\sim 1.4$ mag. Unfortunately, its light curve is poorly sampled to perform a detailed analysis. 
\edit1{The} discovery of SN Refsdal offered us a unique opportunity to model such a distant supernova ($z = 1.5$) and to study properties of its progenitor.  We modelled SN Refsdal using the multi-group radiation-hydrodynamics numerical code \stella which allows one to construct synthetic light curves in various photometric bands and \edit1{accounts} for the expansion velocity of the H$\alpha$ shell. 
For the first time, we obtained the hydrodynamic model of a SN IIP at a cosmologically relevant distance. 
We computed a set of 185 hydrodynamical models covering a rather large area in the parameter space. We confirm the conclusion of \cite{Kelly2016} that SN Refsdal is a more energetic version of SN~1987A.
\edit2{Our calculations suggest that the SN Refsdal progenitor was a blue supergiant star with the radius of  $R_0 = (50 \pm 1) R_{\odot}$, the total mass of $M_{tot}= (25\pm 2) M_{\odot}$, the radioactive \nifsx mass of $M_{^{56}\mathrm{Ni}} = (0.26 \pm 0.05) \,M_{\odot}$, and the total energy release   of $E_{burst}=(4.7 \pm 0.8)\times10^{51}$ erg (parameters were obtained via the Bayesian Model Averaging method).}

%the radius of $R_0 = (50 \pm 1) R_{\odot}$, the total mass of $M_{tot}= (25\pm 2) M_{\odot}$, and the explosion energy of   $E_{burst}=(4.7 \pm 0.8)\times10^{51}$ erg (parameters were obtained via the Bayesian Model Averaging method).}
%a radius of $50 R_{\odot}$, a mass of $26.3 M_{\odot}$ and an explosion energy of $5\times10^{51}$ erg. 
%
%The obtained  ejecta mass and the burst energy are difficult to reconcile with a single star evolution scenario. Instead, the SN Refsdal progenitor could result from a merger of a compact binary system \citep{Menon2017}.
Future deep surveys should be able to detect large number of gravitationally lensed supernovae, including Type IIP, at high redshifts. Analysis of their light curves should allow us to    compare high-$z$ SNe with the local IIP population to investigate any systematic difference between
high and low redshift SNe. 

Proper reconstruction of SN Refsdal light curve allowed us 
\edit1{obtain}
%to tighten existing constraints on 
relative time delays and magnification ratios between images S2-S4 and S1. Mean values (obtained via Bayesian Averaging Method) with uncertainties are provided in Table~\ref{tbl:dtmu}. We anticipated that we would be able to constrain the time delay of the fifth SN Refsdal image SX and its magnification relative to S1 with an accuracy of several percent. Unfortunately, 
quite  a broad `featureless' peak of the light curve 
combined with very scarce photometric measurements for SX  resulted in large uncertainties for parameters of interest. We obtain (again, using BMA method) $\Delta t_{SX-S1}=340^{+43}_{-52}$ days and $\mu_{SX/S1}=0.24^{+0.12}_{-0.07} $.
 Following \edit1{the} approach suggested in \cite{Vega_Ferrero2018}, we computed the Hubble constant $H_0 = 68.6^{+13.6}_{-9.7}$ km s$^{-1}$ Mpc$^{-1}$ via re-scaling the time \edit1{delay} predictions of the lens models to match the values obtained in this work. With more photometric data on SX, \edit1{the} accuracy of $H_0$ determination  can be drastically improved.

\edit1{Not only may next generation telescopes provide  detailed photometric light curves of  lensed SN IIP with resolved images but also high quality spectra which are necessary to determine photospheric velocities at several epochs. With such information, one can significantly improve a pre-supernova model and, as a consequence, obtain a reliable estimate of an absolute magnification of SN images. 
The latter, in its turn, serves as a powerful constraint for lens models. 
Altogether, reliable pre-supernova and lens models ensure robust  and accurate determination of the Hubble constant.
 At the same time,  SNe IIP themselves  with measured photospheric velocities can be used as  direct distance indicators \citep[e.g., ][]{2017hsn..book.2671N}. 
 The same techniques can be applied to lensed SNe IIP if the lens magnification is known. 
 Thus, even a single lensed SNe IIP with spectral information available may in principle provide  two independent probes of $H_0$.  }

%\acknowledgments
\section*{Acknowledgments}

%We thank the anonymous referee for helpful comments to improve the clarity of the paper.
The authors are grateful to Masamune Oguri and Tatiana Larchenkova for many helpful discussions,
Surhud More for valuable comments, which helped to improve the paper.
P.B. is grateful to Ken'ichi Nomoto for the possibility of working at the Kavli IPMU
 and his research has been supported by the grant  RSF 18-12-00522.
S.B. is sponsored by grant RSF 19-12-00229 in his work 
on the supernova simulations with \stella code.
NL acknowledges  support  by grant No. 18-12-00520 from the Russian Scientific Foundation. 
This research has been supported in part by the RFBR (19-52-50014)-JSPS bilateral program.
This work has been supported by World Premier International Research
Center Initiative (WPI), MEXT, Japan, and JSPS KAKENHI Grant Numbers
JP17K05382 and JP20K04024. 
We thank the anonymous referee for useful suggestions and remarks which helped improve the paper.

\software{ \stella, emcee \cite{Foreman-Mackey2012}, matplotlib \citep{Hunter2007}}

% % % % % % % % % % % % % % % % % % % % % % % % % % % %
\appendix
%%\restartappendixnumbering

\section{Fitting model light curves to observations} \label{adx:Bayesian}
Here, we describe our approach of fitting synthetic light curves to observations.
As discussed in Section~\ref{sec:snsim}, we constructed a set of 185 hydrodynamic SN models to find the optimal model which interprets simultaneously available photometric and spectroscopic observations of SN Refsdal. 
Namely, we  use well-sampled measurements  in the F160W, F125W and F105W bands and H$\alpha$ velocities as constraints. 
For each SN model, we derive the  logarithmic likelihood function \eqref{eq:likelihood}:
\begin{equation} \label{eq:likelihood}
%\log p(m_o| m, dt, dm, \sigma)
\log L \left(m^o| m^m, \Theta \right ) = - \frac{1}{2}
% \sum_{Img = [1,4]} \left[
  \sum_{S,t} \left[
%    \sum_{t} \left[
  \frac{\left(m_{S}^{o}(t) - m_{S}^{m}(t+\Delta t) - \Delta m \right)^2}{\sigma_S^2} + \log(2\pi\sigma_S^2)
%     \right] 
  \right]  ,
%\right]
\end{equation}
where $m_S^m(t)$ is the \edit1{model} light curve in a given filter $S$ ($S$ = F160W, F125W or F105W),  
$m_S^o(t)$ is the observed light curve in a filter $S$ sampled at time instances $t$, 
the total  uncertainties $\sigma_S^2 = \sigma_o^2 + \sigma_m^2$ 
are represented with two components: 
the observational photometric uncertainties $\sigma_o$ and the model uncertainties $\sigma_m$. 
Summation is done over three HST filters and time instances at which observations are available.
Vector $\Theta$ denotes the set of five free parameters - the time shift $\Delta t$, \edit1{the magnitude shift} $\Delta m$ and the model uncertainties $\sigma_m$ for F105W, F125W, F160W filters - which we determine for each SN model by  maximizing the log-likelihood~(\ref{eq:likelihood}). 
Since \stella allows one to calculate light curves in multiple bands self-consistently, to match observations we shift all synthetic light curves in brightness  by a single value of $\Delta m$ without adding any filter-related corrections. 
Unfortunately, the `true' model uncertainties are hard to evaluate. 
Moreover, \stella is 1D and makes several simplifying assumptions to numerically treat the radiation hydrodynamics. 
Thus, the perfect fit of a model  to observations does not necessarily leads one to the `true' physical parameters. 
It's more important that the model captures correctly the general shape of the observed LCs. 
That's why we artificially introduced the model uncertainties for each band as fitting parameters. 
Such an approach also allows us to  assign (implicitly) different weights to observations in different bands. 
For example, as can be seen from Table~\ref{tbl:full_param_models} the best-fit model uncertainty for filter  F105W is always larger than $\sigma_m$ for F160W. 
This is partly due to the fact that data in filter F105W are much more sparse and with larger error bars than measurements in F160W.

Having the absolute $\Delta t$ and $\Delta m$ of image S1 fixed, we fit the model light curves to images S2, S3, S4, and  SX by maximizing the function~\eqref{eq:likelihood} with five free parameters:  the time delay of a considered image relative to S1, the magnitude shift relative to S1 and the model uncertainties for three considered HST filters.

\edit1{We use flat priors for all the fit parameters with the following ranges: 
the absolute time shift for the image S1 $ \in $ (-150, 0) days, time shifts for images S2-S4 defined relative to S1  $\in $ (-70, 70) days,  all the magnitude shifts  $\Delta m \in $ (-5, 5),  the model uncertainties $\sigma_m \in $ (0, 1).}
The likelihood distributions are sampled using the  Markov Chain Monte Carlo ensemble sampling tools from the \texttt{emcee} software package \citep{Foreman-Mackey2012}.

Next, we derive the model evidence 
(the marginal distributions of the observations $D$ given the SN model $M_l$ averaged over the prior distributions of all the parameters constituting the vector $\Theta$):
\begin{equation}
\pi(D|M_l) = \sum_i L(\Theta_i) Pr(\Theta_i)
\label{eq:evidence}
\end{equation}

Then we define the posterior probability of each SN model given observations as:
\begin{equation} \label{eq:posterior_prob}
\pi(M_l|D) = \frac{\pi(D|M_l)\pi(M_l)}{\sum_m  \pi(D|M_m) \pi(M_m)}, 
\end{equation}
where we sum up over all 185 SN models in the denominator to ensure that the cumulative posterior probability over all models equals  unity. 
Obtained posterior probabilities $ \pi(M_l|D) $ can be used as a straightforward model selection criteria 
\citep{Hoeting1999}, with the most likely model having the highest value of  $ \pi(M_l|D) $.

% % % % % % % % % % % % % % % % % % % % % % % % % % % % % % % % %
\begin{figure*}[!tbp] 
	\includegraphics[width=\textwidth]{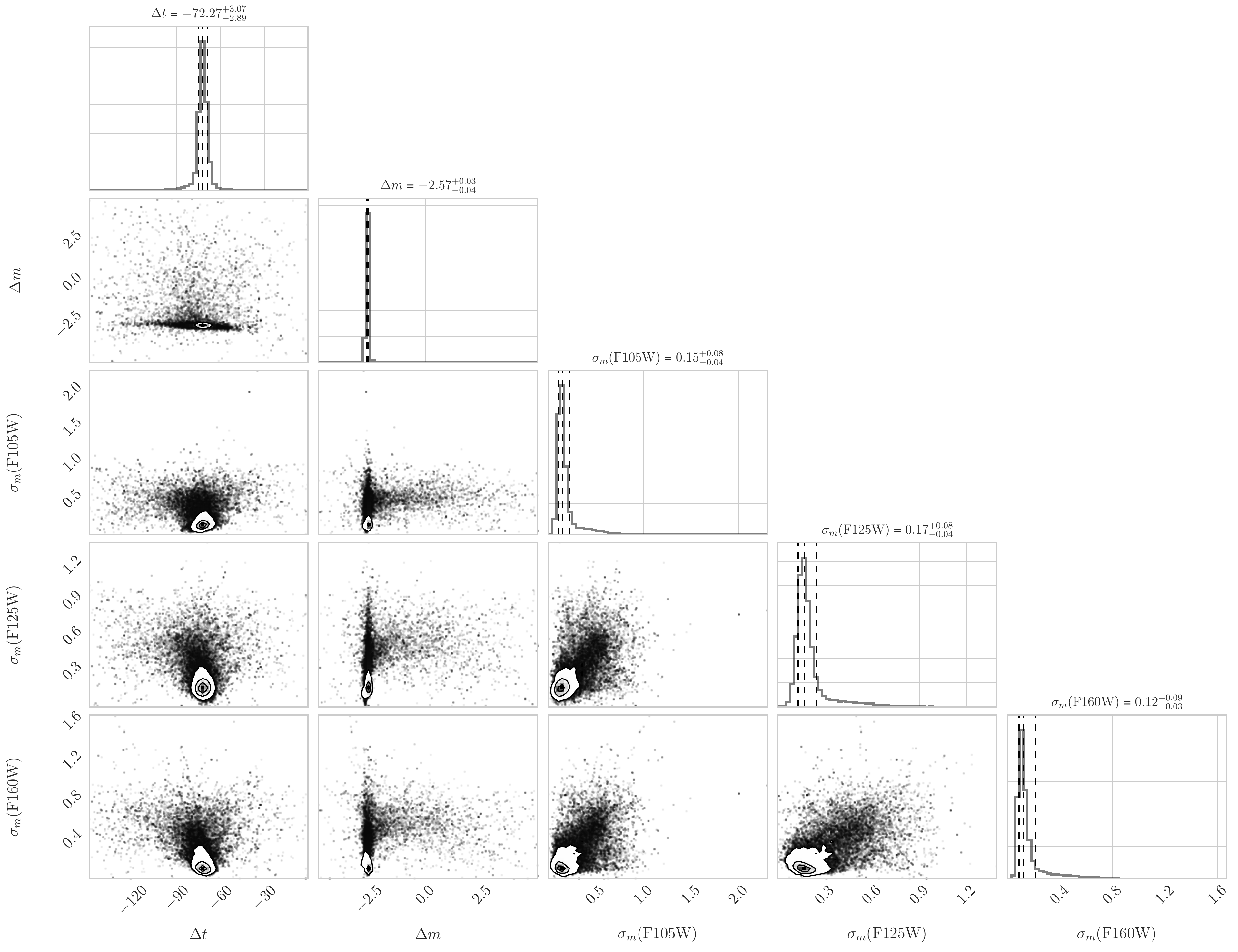}
	\caption{The MCMC corner plot for the best-fit model M1  showing 1D and 2D  marginalized probability contours for each of the five fit parameters (the absolute time shift and \edit1{the absolute magnitude shift} of image S1, and the model uncertainties in three passbands). The histograms on the diagonal also include the positions of the 16th, 50th, and 84th percentiles. }
	\label{fig:corner_refE5R50M26Ni2m2b5m3Z01}
\end{figure*}

\begin{table*}[htb!]  
\centering
	\caption{ 
	Parameters of the 8 best performing SN models}
\begin{tabular}{ccccccc|cccc}
\hline\hline
 Model &  $\pi(M_l|D)$ & $\Delta t$ &  $\Delta m$  &  $\sigma_m(F160W)$ & $\sigma_m(F125W)$ & $\sigma_m(F105W)$ &    $R_0$ &   $M_{tot}$     & $M_{^{56}\mathrm{Ni}}$ & $E_{burst}$  \\
 && (days) &&&	&  &  (\rsun)     & (\msun)         & (\msun)     & (E51)            \\ 
\tableline
M1 & 0.711 & $ {-72}^{+2.6}_{-2.2} $  & $ {-2.57}^{+0.024}_{-0.029} $  & $ {0.11}^{+0.03}_{-0.02} $  & $ {0.16}^{+0.04}_{-0.03} $  & $ {0.14}^{+0.04}_{-0.03} $ & 50.0  & 26.3  & 0.25  & 5.0  \\  % refE5R50M26Ni2m2b5m3Z01 
M2 & 0.165 & $ {-111}^{+2.7}_{-2.7} $  & $ {-2.18}^{+0.027}_{-0.047} $  & $ {0.10}^{+0.05}_{-0.03} $  & $ {0.17}^{+0.05}_{-0.05} $  & $ {0.16}^{+0.05}_{-0.04} $ & 50.0  & 20.6  & 0.37  & 3.0  \\  % refE3R50M20Ni3m2b1m2Z01 
M3 & 0.088 & $ {-71}^{+2.4}_{-2.5} $  & $ {-2.57}^{+0.025}_{-0.028} $  & $ {0.12}^{+0.03}_{-0.02} $  & $ {0.16}^{+0.04}_{-0.03} $  & $ {0.13}^{+0.04}_{-0.03} $ & 50.0  & 26.0  & 0.24 & 5.0  \\  % niCatR50_M26_Ni023_E50 
M4 & 0.0258 & $ {-33}^{+0.4}_{-0.4} $  & $ {-3.17}^{+0.017}_{-0.017} $  & $ {0.10}^{+0.02}_{-0.02} $  & $ {0.17}^{+0.03}_{-0.03} $  & $ {0.11}^{+0.04}_{-0.03} $  & 45.0  & 25.0  & 0.12  & 6.0   \\  % ref_R45_M25_Ni007_E60 
M5 & 0.0047 & $ {-109}^{+2.8}_{-4.5} $  & $ {-2.58}^{+0.018}_{-0.028} $  & $ {0.06}^{+0.03}_{-0.02} $  & $ {0.25}^{+0.05}_{-0.05} $  & $ {0.31}^{+0.07}_{-0.07} $  & 50.0  & 26.3  & 0.24 & 3.0   \\  % refE3R50M26Ni2m2b5m3Z01 
M6 & 0.0022 & $ {-45}^{+3.3}_{-2.4} $  & $ {-3.22}^{+0.020}_{-0.022} $  & $ {0.09}^{+0.02}_{-0.02} $  & $ {0.21}^{+0.04}_{-0.03} $  & $ {0.19}^{+0.05}_{-0.04} $   & 40.0  & 26.0  & 0.12  & 5.5  \\  % ref_R40_M26_Ni007_E55 
M7 & 0.0021 & $ {-87}^{+3.2}_{-3.7} $  & $ {-2.57}^{+0.027}_{-0.062} $  & $ {0.10}^{+0.07}_{-0.03} $  & $ {0.19}^{+0.05}_{-0.06} $  & $ {0.21}^{+0.05}_{-0.06} $  & 50.0  & 26.0  & 0.24  & 4.0  \\  % niCatR50_M26_Ni023_E40 
M8 & 0.0012 & $ {-108}^{+3.3}_{-5.1} $  & $ {-2.59}^{+0.020}_{-0.035} $  & $ {0.07}^{+0.04}_{-0.02} $  & $ {0.25}^{+0.05}_{-0.06} $  & $ {0.31}^{+0.07}_{-0.08} $  & 50.0  & 26.0  & 0.24  & 3.0   \\  % niCatR50_M26_Ni023_E30 
\tableline
\end{tabular} 	\label{tbl:full_param_models}
	\tablecomments{ 
	Parameters of  the best-fit model (with the highest value of $\pi (M_l|D)$ are in the first row). 
	Columns 3-7 list the best values of fit parameters: the absolute time shift of S1 in days relative \edit1{to} MJD$_0=57000$, the absolute magnitude shift of S1, and the model uncertainties in three HST passbands.
	Columns 8-11 give information on physical parameters of \edit1{each} SN model, namely, \edit1{the} pre-supernova radius and \edit1{the total} mass, \edit1{the} radioactive \nifsx mass and \edit1{the} explosion energy (in units of $10^{51}$ erg). }
\end{table*}

For each SN model in our set,  we evaluate  $ \pi(M_l|D) $ by comparing the synthetic light curves with observations for SN Refsdal images S1-S4 in F105W, F125W, F160W pass bands. 
As a result, for each SN model we obtain $ \pi(M_l|D) $ and the best-fitting parameters:  the absolute time shift of image S1, i.e. modified Julian date of the explosion (MJD$_{exp}$), and the absolute magnification of S1 (both of which are actually nuisance parameters), the time shifts of images S2-S4 relative to S1 and magnification ratios. Table~\ref{tbl:full_param_models} lists the best-fit parameters for S1 (the absolute time shift $\Delta t$, the absolute magnitude shift $\Delta m$, and the model uncertainties $\sigma_m$ for F105W, F125W, F160W filters) as well as the basic SN model characteristics for 8 best performing models. Note that the best-fit  absolute magnifications of S1 $\mu = 10^{-\Delta m/2.5}\simeq 7.5 \div 20$ for SN models in  Table~\ref{tbl:full_param_models}  \edit2{are in  approximate agreement with }   absolute magnifications predicted by lens models \citep{Oguri2015, Kawamata2016, Sharon2015, Jauzac2015, Grillo2016}. This is not a result of fine-tuning since for $\Delta m$ determination we used a flat prior in a wide range of values $\Delta m \in $ (-5, 5). 
Table~\ref{tbl:models_dt_dm} shows relative time delays for images S2-S4 in days, magnification ratios as well as the explosion and peak MJDs. Despite the fact that for the top three SN models (the first three rows in Table~\ref{tbl:models_dt_dm}) the explosion dates (or the absolute time shift in Table~\ref{tbl:full_param_models}) vary noticeably, the resulting relative time delays are not that different. \edit1{A} similar conclusion can be made about magnifications. The absolute magnification varies at most by a factor of $\sim 2.5$, while the relative magnifications (columns 9-11 in Table~\ref{tbl:models_dt_dm}) show variations by several percent only.
Table~\ref{tbl:full_param_models} also provides values of $ \pi(M_l|D) $ for each model which reflect quality of fit to the data. The best-fit model M1 (see Table~\ref{tbl:full_param_models})  significantly outperforms all the others (what is not surprising since it was our goal to construct the SN model which matches  best available SN Refsdal observations).  For the best-fit model,  Figure~\ref{fig:corner_refE5R50M26Ni2m2b5m3Z01} plots 
2D and 1D probability distributions of each of the five fit parameters.

\begin{table*}[htb!]  
\centering
	\caption{Time delay and magnification measurements for SN Refsdal images S1-S4.}
\begin{tabular}{ccccccccccc}
\hline\hline
Model  &  $\pi(M_l|D)$  &  $\text{MJD}_{exp}$  &  $\text{MJD}_{pk}$  &  $\Delta t_{S2-S1}$  &  $\Delta t_{S3-S1}$  &  $\Delta t_{S4-S1}$  &  $\mu_{S1}$  &  $\mu_{S2/S1}$  &  $\mu_{S3/S1}$  &  $\mu_{S4/S1}$  \\
\tableline 
$ M1 $   &   $ 0.711 $   &   $ 56928^{+2.57}_{-2.23} $   &   $ 57145^{+2.6}_{-2.2} $   &   $ 10.0^{+1.93}_{-2.04} $   &   $ 4.2^{+2.35}_{-2.34} $   &   $ 30.4^{+6.59}_{-8.00} $   &   $ 10.62^{+0.285}_{-0.229} $   &   $ 1.14^{+0.019}_{-0.019} $   &   $ 1.01^{+0.019}_{-0.018} $   &   $ 0.34^{+0.015}_{-0.014} $  \\  % refE5R50M26Ni2m2b5m3Z01  
 $ M2 $   &   $ 0.165 $   &   $ 56889^{+2.70}_{-2.79} $   &   $ 57141^{+2.7}_{-2.8} $   &   $ 8.7^{+2.43}_{-2.45} $   &   $ 4.5^{+1.82}_{-1.97} $   &   $ 31.0^{+9.28}_{-7.81} $   &   $ 7.43^{+0.327}_{-0.176} $   &   $ 1.12^{+0.019}_{-0.019} $   &   $ 1.01^{+0.015}_{-0.014} $   &   $ 0.35^{+0.019}_{-0.015} $  \\  % refE3R50M20Ni3m2b1m2Z01  
 $ M3 $   &   $ 0.088 $   &   $ 56929^{+2.37}_{-2.47} $   &   $ 57138^{+2.4}_{-2.5} $   &   $ 11.2^{+1.86}_{-2.44} $   &   $ 4.7^{+2.62}_{-2.54} $   &   $ 28.7^{+6.70}_{-4.62} $   &   $ 10.71^{+0.283}_{-0.248} $   &   $ 1.14^{+0.020}_{-0.020} $   &   $ 1.01^{+0.020}_{-0.019} $   &   $ 0.34^{+0.013}_{-0.012} $  \\  % niCatR50_M26_Ni023_E50  
 $ M4 $   &   $ 0.0258 $   &   $ 56967^{+0.37}_{-0.45} $   &   $ 57134^{+0.4}_{-0.4} $   &   $ 0.5^{+0.28}_{-0.38} $   &   $ 0.5^{+0.28}_{-0.40} $   &   $ 12.2^{+1.04}_{-1.91} $   &   $ 18.59^{+0.302}_{-0.292} $   &   $ 1.11^{+0.018}_{-0.019} $   &   $ 1.01^{+0.016}_{-0.016} $   &   $ 0.33^{+0.010}_{-0.010} $  \\  % ref_R45_M25_Ni007_E60  
 $ M5 $   &   $ 0.0047 $   &   $ 56891^{+2.95}_{-5.49} $   &   $ 57150^{+3.0}_{-5.5} $   &   $ 9.4^{+3.20}_{-3.61} $   &   $ 4.7^{+1.85}_{-1.93} $   &   $ 34.0^{+10.84}_{-10.04} $   &   $ 10.74^{+0.403}_{-0.186} $   &   $ 1.15^{+0.024}_{-0.022} $   &   $ 1.03^{+0.013}_{-0.013} $   &   $ 0.38^{+0.024}_{-0.019} $  \\  % refE3R50M26Ni2m2b5m3Z01  
 $ M6 $   &   $ 0.0022 $   &   $ 56955^{+3.26}_{-2.48} $   &   $ 57166^{+3.3}_{-2.5} $   &   $ 12.0^{+0.54}_{-0.92} $   &   $ 1.0^{+1.38}_{-1.83} $   &   $ 20.5^{+3.32}_{-5.66} $   &   $ 19.41^{+0.388}_{-0.345} $   &   $ 1.15^{+0.019}_{-0.018} $   &   $ 1.00^{+0.013}_{-0.013} $   &   $ 0.35^{+0.013}_{-0.013} $  \\  % ref_R40_M26_Ni007_E55  
 $ M7 $   &   $ 0.0021$   &   $ 56913^{+3.23}_{-3.72} $   &   $ 57152^{+3.2}_{-3.7} $   &   $ 10.7^{+2.85}_{-3.23} $   &   $ 5.3^{+2.62}_{-2.61} $   &   $ 33.1^{+7.55}_{-7.27} $   &   $ 10.69^{+0.648}_{-0.265} $   &   $ 1.14^{+0.022}_{-0.021} $   &   $ 1.02^{+0.019}_{-0.017} $   &   $ 0.36^{+0.016}_{-0.015} $  \\  % niCatR50_M26_Ni023_E40  
 $ M8 $   &   $ 0.0012 $   &   $ 56892^{+3.12}_{-4.66} $   &   $ 57145^{+3.1}_{-4.7} $   &   $ 8.8^{+3.30}_{-3.34} $   &   $ 4.9^{+2.26}_{-2.34} $   &   $ 32.4^{+11.35}_{-10.21} $   &   $ 10.81^{+0.344}_{-0.192} $   &   $ 1.15^{+0.023}_{-0.022} $   &   $ 1.03^{+0.014}_{-0.014} $   &   $ 0.37^{+0.024}_{-0.018} $  \\  % niCatR50_M26_Ni023_E30  
\tableline
 BMA    &  ---  &   $ 56922^{+18.6}_{-18.6} $   &   $ 57144^{+10}_{-10} $   &   $ 9.5^{+2.6}_{-2.7}  $   &   $ 4.2^{+2.3}_{-2.3} $   &   $ 30^{+7.8}_{-8.2} $   &   $ 10.2^{+1.9}_{-1.6} $   &   $ 1.14^{+0.021}_{-0.020} $   &   $ 1.01^{+0.019}_{-0.018} $   &   $ 0.35^{+0.016}_{-0.015} $  \\  % refE5R70M26Ni2m2b5m3Z01h5  
\tableline
\end{tabular} 	\label{tbl:models_dt_dm}
\tablecomments{ Columns 3 and 4 show the modified Julian dates of the explosion and the peak, correspondingly. Columns 5-7 provide the time delays (in days) relative to S1. Column 8 lists the absolute magnifications for S1, while columns 9-11 give magnifications of images S2-S4 relative to S1. }
\end{table*}

\begin{figure*}  
\centering
	\includegraphics[width=0.8\textwidth]{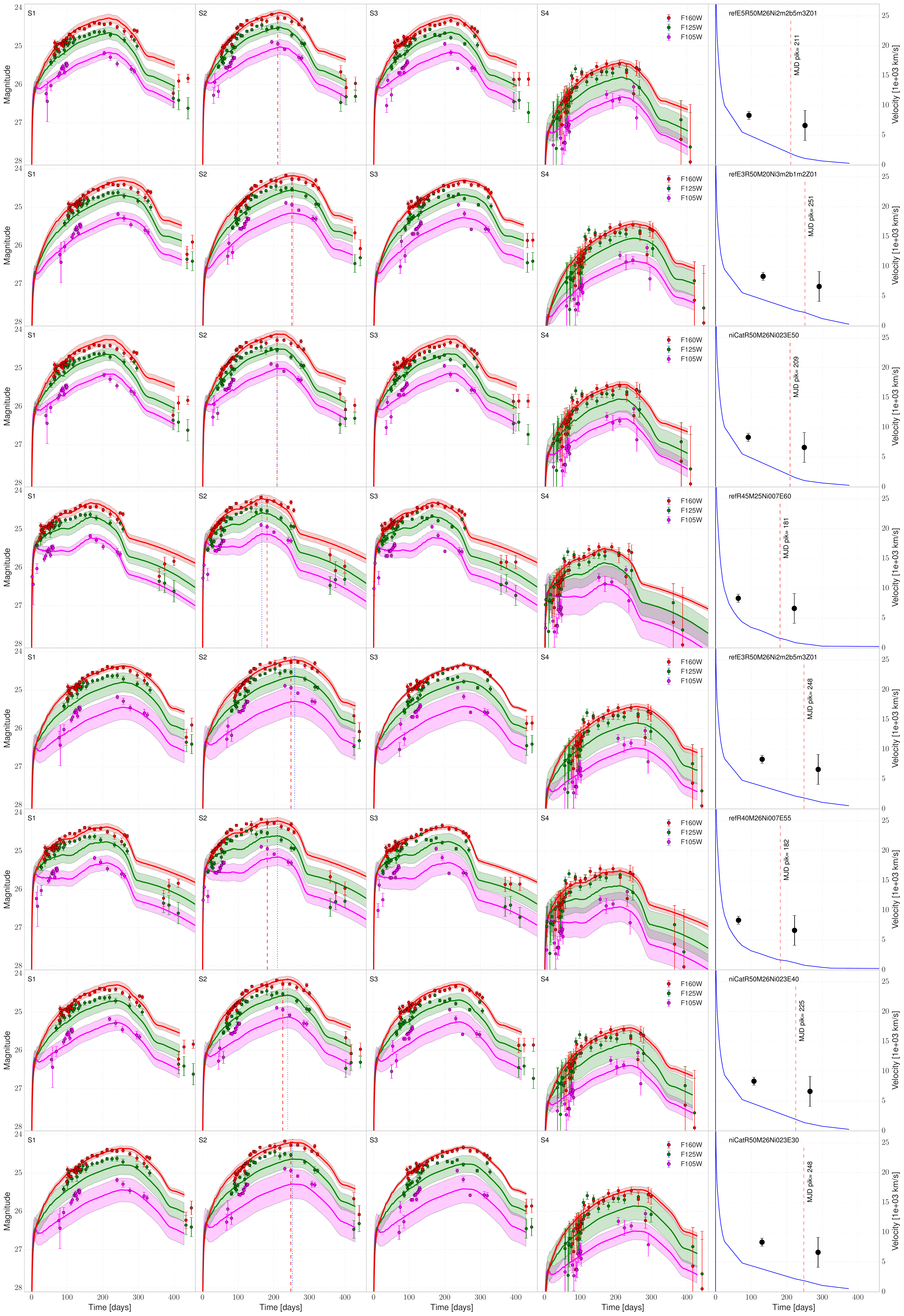}
	\caption{ Light curves and the photosphere velocity evolution of the best-performing SN models listed in Table~\ref{tbl:full_param_models}. The best-fit model M1  is shown in the first row, M2 is in the second, ..., M8 is in the eighth row. In each row, the first four panels show the observed and model LCs for SN Refsdal images S1-S4 (from left to right). Shaded areas indicate the model uncertainties in each passband. The most right panel in each row plots the model photospheric velocity evolution in comparison with H$\alpha$-velocity measurements from \citealt{Kelly2016}. As expected, H$\alpha$-velocities are systematically higher than the photospheric velocities. \edit1{The vertical dashed line shows the date of peak brightness of a model light curve in F160W passband.} }
	\label{fig:ubv_obs_models_S1S4}
\end{figure*}

We compute a weighted average and an uncertainty of parameters of interest $\Theta$ across the explored set of SN models $M_l$ using the Bayesian Model Averaging  approach \citep{Hoeting1999}:
\begin{equation}
  E[\Theta]   = \sum_l \pi(M_l|D) \;  \Theta_l 
  \label{eq:bma1}  
\end{equation}
\begin{equation}
Var[\Theta] = \left|\sum_l \pi(M_l|D) \left(  Var(\Theta_l) + \Theta^2 \right)- E[\Theta]^2 \right |
\label{eq:bma2}
\end{equation}

\begin{figure*}
\includegraphics[width=\linewidth]{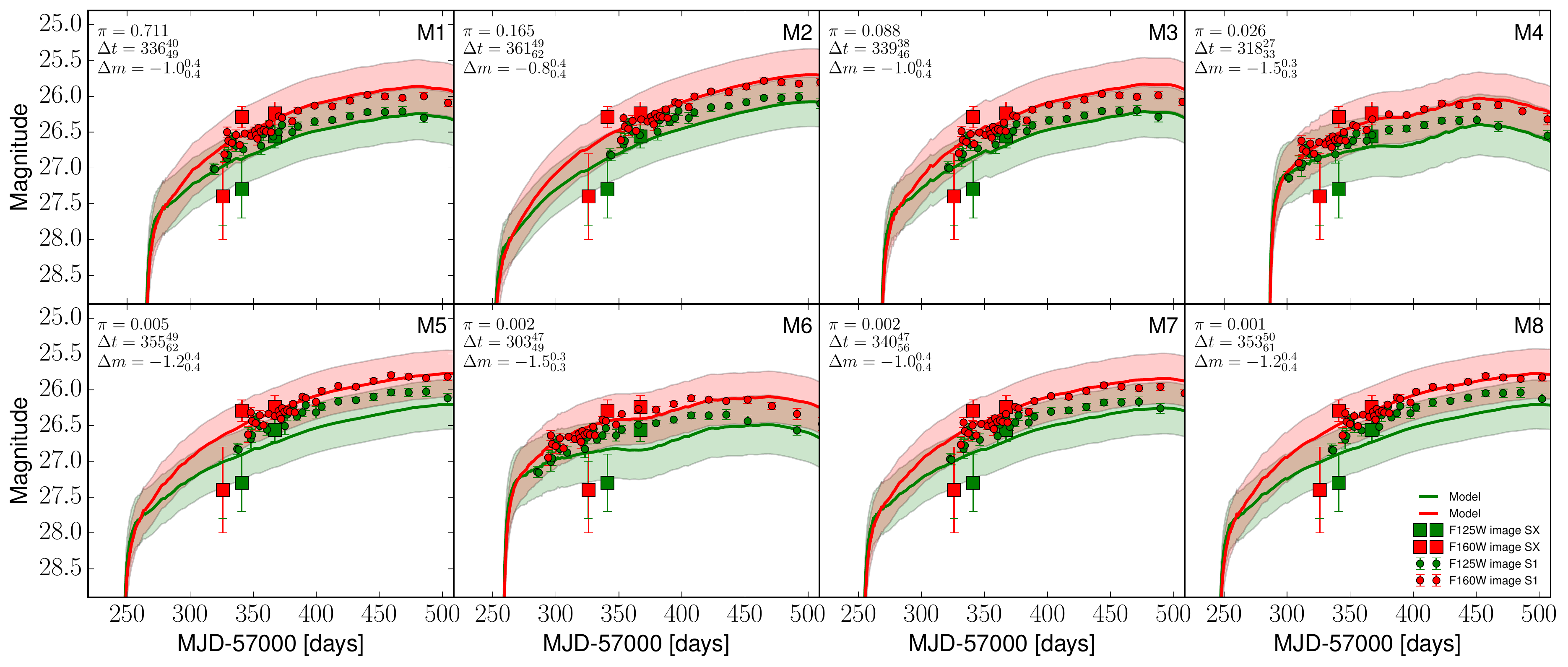}
\caption{The model light curves of 8 best performing models fitted to the SX data (shows as squares). For comparison, S1 data are overplotted as circles.  } 
\label{fig:ubv_SX_fit_models_t0}
\end{figure*}

\begin{figure*}
\includegraphics[width=\linewidth]{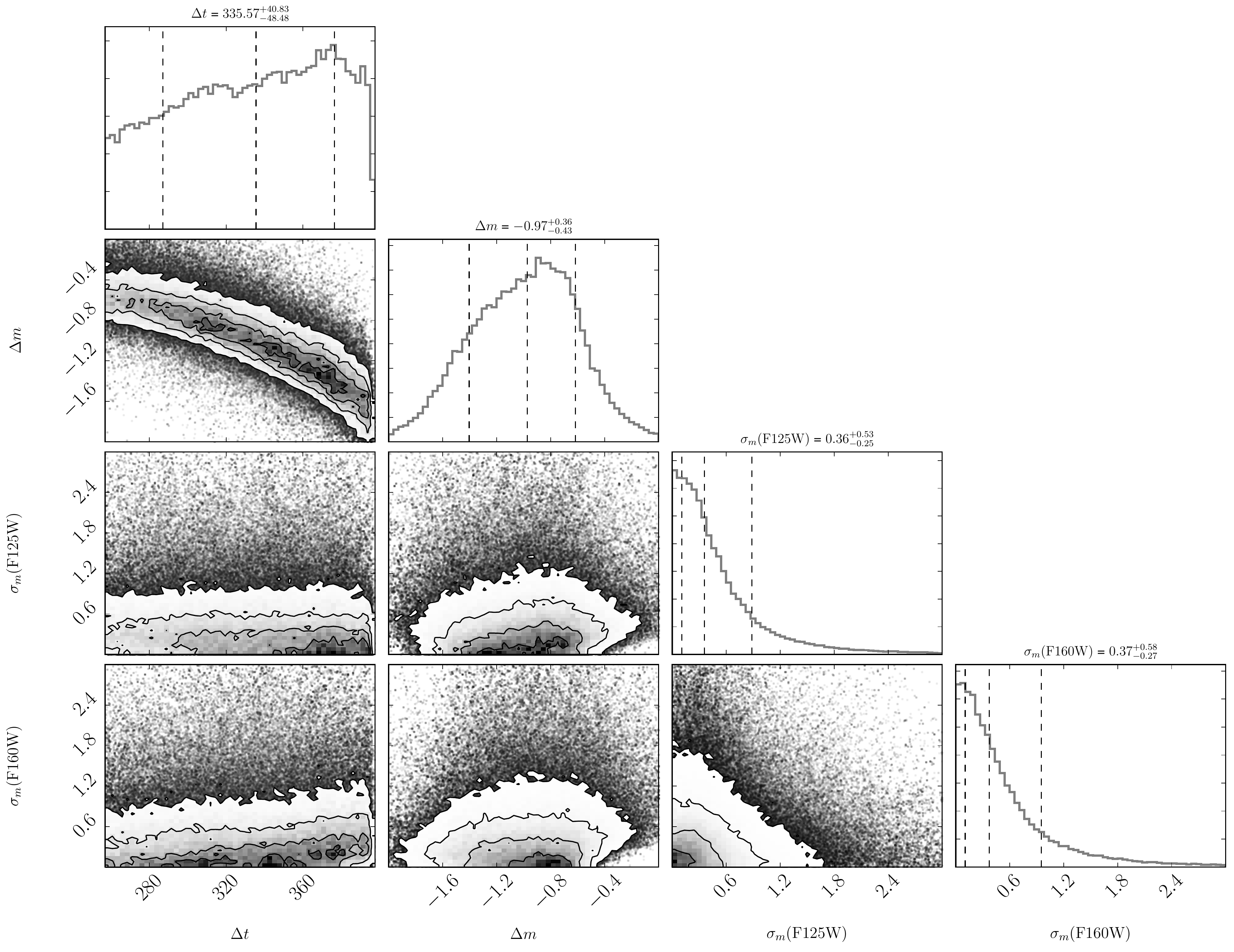}
\caption{The MCMC corner plot for the best-fit model M1  showing marginal distributions for the following fit parameters: the SX-S1 time delay, the absolute magnitude shift for SX  (which can be related to the magnification ratio as $\mu_{SX/S1} = 10^{-dm/2.5}/\mu_{S1}$, where $\mu_{S1}$=10.62 is the absolute magnification for S1) and the model uncertainties in F125W and F160W filters. Dashed vertical lines in the histograms on the diagonal mark the positions of the 16th, 50th, and 84th percentiles. } 
\label{fig:triangle_all}
\end{figure*}

The same procedure of finding the best-fit time and magnitudes shits is applied to SN Refsdal image SX. 
\edit1{We  use flat priors on free parameters: the absolute magnitude shift $\Delta m \in (-2,0)$, model uncertainties $\sigma_m \in (0,1)$. The time delay relative to S1 is varied within the range which depends on a SN model and is defined from the following two conditions: (i) in January 2016 SX showed brightening \citep{Kelly2016_SNX}, and (ii) SX appeared no earlier than on 2005 October 30 \citep{Kelly2016_SNX}.  }

The best performing models are shown in Figure~\ref{fig:ubv_SX_fit_models_t0}. The marginal distributions of fit parameters for the best-fit model M1 are illustrated in Figure~\ref{fig:triangle_all}. The final time delay and magnification ratio estimates for SX (relative to S1) are obtained again via Bayesian Model Averaging and provided in Section~\ref{subsec:SX}.

% % % % % % % % % % % % % % % % % % % % % % % % % % % % % % % % %
\bibliographystyle{aasjournal} %  aasjournal myjetpl, plain, abbrv, unsrt, alpha
\bibliography{refs2}

\listofchanges

\end{document}